\documentclass{imamat}
\usepackage{graphicx}
\usepackage{subfig}

\jno{drnxxx}

\usepackage{color}
\usepackage{amssymb}

\def\spacce#1{\hskip #1pt}
\def\drawline#1#2{\raise 2.5pt\vbox{\hrule width #1pt height #2pt}}

\def\solid{\drawline{22}{1.0}\nobreak\ }

\def\tdash{\hbox{\drawline{4}{1.0}\spacce{2}}}
\def\dashed{\tdash \tdash \tdash \tdash \nobreak\ }

\def\tdot{\hbox{\drawline{1}{1.0}\spacce{2}}}
\def\dotted{\tdot \tdot \tdot \tdot \tdot \tdot \tdot \tdot \nobreak\ }

\def\dashdot{\tdash \tdot \tdash \tdot \tdash \nobreak\ }

\def\ssquare{${\vcenter{\hrule height 0.9pt
       \hbox{\vrule width 0.9pt height 10pt \kern 10pt
       \vrule width 0.9pt}
       \hrule height 0.9pt}}$\nobreak\ }
       
\def\ssquareb{$\Box$\nobreak\ }       
       
\def\blackssquare{$\scriptstyle\blacksquare$\nobreak\ }
       
\def\circle{$\circ$\nobreak\ }

\def\blackcircle{$\bullet$\nobreak\ }

\def\bigcircle{$\bigcirc$\nobreak\ }

\def\losange{$\Diamond$\nobreak\ }

\def\blacklosange{$\blacklozenge$\nobreak\ }

\def\trianup{\raise 1.25pt\hbox{$\triangle$}\nobreak\ }

\def\blacktrianup{\raise 1.25pt\hbox{$\blacktriangle$}\nobreak\ }

\def\bigtriandown{\raise 1.25pt\hbox{$\bigtriangledown$}\nobreak\ } 

\def\plus{$+$ \nobreak}

\def\asterix{$\ast$ \nobreak}

\def\solidopencircle{\solid \nobreak\spacce{-19.5}\raise
 -0.5pt\hbox{\circle} \nobreak\ }
 
\def\solidblackcircle{\solid \nobreak\spacce{-21}\raise
 -0pt\hbox{\blackcircle} \nobreak\ }

\def\solidopensquare{\solid \nobreak\spacce{-19}\raise
 -1.0pt\hbox{\ssquare} \nobreak\ }
 
\def\solidblacksquare{\solid \nobreak\spacce{-19}\raise
 -1.0pt\hbox{\blackssquare} \nobreak\ }
 
\def\solidopenlosange{\solid \nobreak\spacce{-19}\raise
 -1.0pt\hbox{\losange} \nobreak\ }
 
\def\solidblacklosange{\solid \nobreak\spacce{-21}\raise
 -2.0pt\hbox{\blacklosange} \nobreak\ }

\def\solidopentriangleup{\solid \nobreak\spacce{-19}\raise
 -1.pt\hbox{\trianup} \nobreak\ }
 
\def\solidblacktriangleup{\solid \nobreak\spacce{-21.5}\raise
 -2.0pt\hbox{\blacktrianup} \nobreak\ }

\def\solidopencross{\solid \nobreak\spacce{-22}\raise
 -1.0pt\hbox{\cross} \nobreak\ }
 
\def\solidopenplus{\solid \nobreak\spacce{-21}\raise
 -1.5pt\hbox{\plus} \nobreak\ }
 
\def\solidopenasterix{\solid \nobreak\spacce{-21}\raise
 -1.5pt\hbox{\asterix} \nobreak\ }

\def\dashedopencircle{\dashed \nobreak\spacce{-22}\raise
 -1.0pt\hbox{\circle} \nobreak\ }
 
\def\dashedblackcircle{\dashed \nobreak\spacce{-23}\raise
 -1.4pt\hbox{\blackcircle} \nobreak\ }
 
\def\dashedopensquare{\dashed \nobreak\spacce{-22}\raise
 -1.0pt\hbox{\ssquare} \nobreak\ }
 
\def\dashedblacksquare{\dashed \nobreak\spacce{-23}\raise
 -1.0pt\hbox{\blackssquare} \nobreak\ }
 
\def\dashedopenlosange{\dashed \nobreak\spacce{-22}\raise
 -1.0pt\hbox{\losange} \nobreak\ }
 
\def\dashedblacklosange{\dashed \nobreak\spacce{-23}\raise
 -2.0pt\hbox{\blacklosange} \nobreak\ }
 
\def\dashedopentriangleup{\dashed \nobreak\spacce{-22}\raise
 -1.0pt\hbox{\trianup} \nobreak\ }
 
\def\dashedblacktriangleup{\dashed \nobreak\spacce{-24}\raise
 -2.0pt\hbox{\blacktrianup} \nobreak\ }
 
\def\dashedopencross{\dashed \nobreak\spacce{-24}\raise
 -1.0pt\hbox{\cross} \nobreak\ }
 
\def\dashedopenplus{\dashed \nobreak\spacce{-24}\raise
 -1.0pt\hbox{\plus} \nobreak\ }

\def\dashedopenasterix{\dashed \nobreak\spacce{-22}\raise
 -1.5pt\hbox{\asterix} \nobreak\ }

\def\dottedopencircle{\dotted \nobreak\spacce{-22}\raise
 -1.0pt\hbox{\circle} \nobreak\ }
 
\def\dottedblackcircle{\dotted \nobreak\spacce{-23}\raise
 -1.4pt\hbox{\blackcircle} \nobreak\ }

\def\dottedopensquare{\dotted \nobreak\spacce{-22}\raise
 -1.0pt\hbox{\ssquare} \nobreak\ }
 
\def\dotted\mathbfuare{\dotted \nobreak\spacce{-23}\raise
 -1.0pt\hbox{\blackssquare} \nobreak\ }
 
\def\dottedopenlosange{\dotted \nobreak\spacce{-22}\raise
 -1.0pt\hbox{\losange} \nobreak\ }
 
\def\dottedblacklosange{\dotted \nobreak\spacce{-24}\raise
 -2.0pt\hbox{\blacklosange} \nobreak\ }
 
\def\dottedopentriangleup{\dotted \nobreak\spacce{-22}\raise
 -1.0pt\hbox{\trianup} \nobreak\ }
 
\def\dottedblacktriangleup{\dotted \nobreak\spacce{-25}\raise
 -2.0pt\hbox{\blacktrianup} \nobreak\ }
 
\def\dottedopencross{\dotted \nobreak\spacce{-23.5}\raise
 -1.0pt\hbox{\cross} \nobreak\ }
 
\def\dottedopenplus{\dotted \nobreak\spacce{-23.5}\raise
 -1.0pt\hbox{\plus} \nobreak\ }

\def\dottedopenasterix{\dotted \nobreak\spacce{-23.5}\raise
 -1.0pt\hbox{\asterix} \nobreak\ }

\def\whitehisto{${\vcenter{\color{black} \hrule height 1.0pt
       \hbox{\vrule width 1.0pt height 4pt \kern 8pt
       \vrule width 1.0pt}
       \hrule height 1.0pt}}$\nobreak\ }
       
\def\histosymb#1{${\vcenter{\color{#1} \hrule height 0.0pt
       \hbox{\vrule width 11.0pt height 6pt \kern 0pt 
       \vrule width 0.0pt}
       \hrule height 0.0pt}}$\nobreak\ }

\def\symbol#1#2#3#4#5#6{\color{#5}#1 \nobreak\spacce{-#3}\raise
 -#4pt\hbox{\color{#6}#2}\color{black} \nobreak\ }

\begin{document}

\title{Simulations of Brownian tracer transport in squirmer suspensions}
\shorttitle{Brownian tracer transport in cell suspensions}

\author{%
{\sc
Blaise Delmotte \thanks{ Email: delmotte@cims.nyu.edu}}\\[2pt]
Courant Institute of Mathematical Sciences, New York University, New York, NY 10012, USA\\[6pt]
{\sc and}\\[6pt]
{\sc Eric E Keaveny\thanks{Email: e.keaveny@imperial.ac.uk}}\\[2pt]
Department of Mathematics, Imperial College London,\\
South Kensington Campus, London, SW7 2AZ,UK\\[6pt]
{\sc and}\\[6pt]
{\sc Eric Climent \thanks{Email: ecliment@imft.fr} and Franck Plourabou\'e \thanks{Email: fplourab@imft.fr} }\\[2pt]
Institut de M\'{e}canique des Fluides de Toulouse (IMFT) - Universit\'e de Toulouse,\\ CNRS-INPT-UPS, FR-31400 Toulouse, France
}
\shortauthorlist{B. Delmotte \emph{et al.}}

\maketitle

\begin{abstract}
{
In addition to enabling movement towards environments with favourable living conditions, swimming by microorganisms has also been linked to enhanced mixing and improved nutrient uptake by their populations.  Experimental studies have shown that Brownian tracer particles exhibit enhanced diffusion due to the swimmers, while theoretical models have linked this increase in diffusion to the flows generated by the swimming microorganisms, as well as collisions with the swimmers.  In this study, we perform detailed simulations based on the force-coupling method and its recent extensions to the swimming and Brownian particles to examine tracer displacements and effective tracer diffusivity in squirmer suspensions.  By isolating effects such as hydrodynamic or steric interactions, we provide physical insight into experimental measurements of the tracer displacement distribution.  In addition, we extend results to the semi-dilute regime where the swimmer-swimmer interactions affect tracer transport and the effective tracer diffusivity no longer scales linearly with the swimmer volume fraction.
}
{Tracer dispersion - Squirmers - Active suspensions - Simulations}
\end{abstract}

\section{Introduction}
Active suspensions of micro-swimmers such as spermatozoa (\cite{Creppy2015}), bacteria (\cite{Wensink2012}) or microalgae are common both in the natural environment, such as oceans (\cite{Pedley1992,Stocker2012,Durham2013}), lakes, ponds  and within living organisms, such as the human body.  Such suspensions can exhibit the formation of coherent structures or complex flow patterns (\cite{Saintillan2011,Marchetti2013,Wensink2012}) which may lead to enhanced mixing of chemicals in the surrounding fluid, the alteration of suspension rheology (\cite{Rafai2010,Lopez2015}), or increased nutrient uptake. Mixing and transport of microscopic, inert particles by motile microorganisms have been a topic of recent interest as such suspensions are a prime example of out-of-equilibrium systems.  The particles experience Brownian motion due to their small size and are further affected by hydrodynamics interactions and collisions with the micro-swimmers.  Enhanced particle transport in active suspensions has been observed in the presence of collective motion (\cite{Wu2000}), but also in the absence of it (\cite{Kurtuldu2011}).  Understanding the mechanism of enhanced tracer transport can provide insight into biological processes such as predator-prey interactions and the plankton food chain (\cite{Kiorboe2014}), as well as chemical signalling and quorum-sensing (\cite{Kim2016}). In addition, the underlying mechanism may also provide the foundation for the design of novel biomimetic micro-fluidic devices that use similar strategies for enhanced mixing and stirring at small scales. 

Generally speaking, the motion of a small particle in a suspension of micro-swimmers results from the interplay between Brownian diffusion due to thermal fluctuations and transport by the flow field induced by all the swimmers.  The diffusion coefficient of the tracer, $D_0$ is set by the Stokes-Einstein diffusivity 
\begin{equation}
 D_0 = \dfrac{k_BT}{\zeta},
 \label{eq:Stokes_einstein}
\end{equation}
where $\zeta$ is the friction coefficient of the tracer particle and $k_BT$ the thermal energy.  The tracer particles are advected by the velocity field $u(\mathbf{x})$ which corresponds to the disturbances generated by the collection of micro-swimmers in the vicinity of the tracer.  The velocity $u(\mathbf{x})$ depends highly on the magnitude and the spatial rate of decay of the disturbances, as well as the local swimmer volume fraction, $\phi_v$.  

As the swimmers are moving, over time, the ever changing advection of the tracer particles results in their diffusive behaviour.  In the dilute regime where $\phi_v \ll 1$, many experiments (\cite{Leptos2009,Mino2011,Jepson2013,Kasyap2014}) have shown that the scaling between the tracer diffusivity, $D^{act}$, due to swimmer activity and swimmer volume fraction is linear
\begin{equation}
  D^{act} = D_{eff}-D_0 = \alpha \phi_v
  \label{eq:linear_diff}
\end{equation}
where  $D_{eff}$ is the total effective tracer diffusivity and $\alpha$ has the units of a diffusion coefficient.  In terms of swimmer number density, $n$ and swimming speed, $U$, this can also be expressed as $D^{act} = nU\Lambda$, where $nU$ is the ``active flux'' of micro-swimmers and $\Lambda$ scales like the fourth power of the swimmer size based on dimensional analysis.  In the dilute limit, where swimmers move along straight paths and the interactions between tracers and swimmers are well characterised, the effective diffusion coefficient can be computed (\cite{Lin2011, Mino2011,Pushkin2013,Pushkin2013b, Kasyap2014,Thiffeault2014, Burkholder2017}) by averaging the displacements of a single particle due to repeated interactions with swimmers that move independently of one another.  In many of these studies, the swimmers are modelled as spherical squirmers. The spherical squirmer model was first introduced by \cite{Lighthill1952} and furthered by \cite{Blake1971} and provides the motion and flow generated by a spherical swimmer propelled by small distortions of its surface.

Theoretical studies have shown that $D^{act}$ can be separated into two contributions: random swimmer reorientations and the entrainment by the swimmer along the straight trajectories.  In the experiments of \cite{Leptos2009}, the tracer diffusion coefficient in 3D dilute suspensions of \emph{C. Rheinardtii} was obtained from data taken over period of two seconds.  Since the reorientation time of \emph{C. Rheinardtii} due to phase slips between flagella is approximately ten seconds (\cite{Goldstein2009}), swimmer reorientation does not significantly affect the tracer diffusion.  Hence, by considering entrainment only, \cite{Pushkin2013} obtained an estimate ($(D_{eff}-D_0)/\phi_v \simeq 83\mu m^2s^{-1}$), surprisingly close to the experiments of \cite{Leptos2009} ($(D_{eff}-D_0)/\phi_v \simeq 81.3\mu m^2s^{-1}$).  Using a similar approach, \cite{Thiffeault2014} showed that the distribution of tracer displacements from \cite{Leptos2009} can also be reproduced using the squirmer model.  However, these two works only consider point tracers while particles in experiments are micron-sized beads (\cite{Wu2000,Leptos2009,Mino2011,Kasyap2014}), macromolecules, or dead cells (\cite{Jepson2013}) with a finite size.  Recent experiments (\cite{Polin2016}) in the dilute regime have shown that the front-mounted flagella of \emph{C. Rheinardtii} can trap such micron-sized particles and generate very large displacements through direct entrainment.  Such dramatic events are due to near-field interactions and are strongly related to swimmer geometry, as well as actuation mechanism.

Beyond the dilute limit, there can be a break down in the linear dependence of the effective diffusion coefficient on swimmer volume fraction.  The departure from linearity, however, appears to depend strongly on the details of the system used for study.  For example, \cite{Wu2000} showed that the linear scaling holds up to $\phi_v = 10\%$ in two-dimensional films of \emph{E. coli}, while \cite{Kasyap2014} observed in a three-dimensional bath that the linear trend breaks down for $\phi_v\geq 2.5\%$.  As stated by \cite{Kasyap2014}, the reason for the deviation from linearity is still unclear, and ``one possibility is the occurrence of multi-bacterial effects on the tracer diffusivity."  While suspensions of bacteria are known to display large-scale collective dynamics (\cite{Wu2000,dombrowski_04}) characterised by swirls and jets, the nonlinear dependence is also observe in swimmer suspensions that do not exhibit larger scale motion.  For example, \cite{Kurtuldu2011} measured tracer displacements in a suspension of \emph{C. Rheinardtii} confined to a thin film of liquid and obtained a power-law for the effective diffusivity $D_{eff}/D_0 \sim \phi_v^{3/2}$.   A numerical investigation of the semi-dilute regime using the squirmer model and Stokesian Dynamics \cite{Ishikawa2010}) to compute the motion of non-Brownian fluid particles yielded a linear scaling of the tracer diffusivity with volume fraction for values up to $\phi_v = 15\%$.  

In this paper, we present results from simulations exploring tracer transport in dilute and semi-dilute suspensions of squirmers.  In our simulations, we employ recently developed numerical tools based on the force-coupling method (FCM), (\cite{Maxey2001,Lomholt2003,Keaveny2014,Delmotte_2015a,Delmotte_2015b}) that allow for multibody hydrodynamic interactions between active and passive particles, polydispersity in particle size, particle Brownian motion that satisfies the fluctuation-dissipation theorem, and steric interactions.   A description of the squirmer model and fluctuating FCM are provided in Section \ref{sec:model}.  In the dilute regime, we obtain quantitative agreement between our simulation results and the experimental results of \cite{Leptos2009} for the effective tracer diffusion coefficient, as well as for the tracer displacement distribution.  By selectively removing the flow disturbances due to the squirming modes, we quantify the contributions of hydrodynamic and steric interactions on tracer displacement and provide insight into physical mechanisms giving rise to particular features of the tracer displacement distribution.  These results are shown in Section \ref{sec:Dilute}.  We extend these results to semi-dilute concentrations in Section \ref{sec:tracer_concentrated}.  Here, we examine the non-linear dependence of the effective tracer diffusion coefficient on the swimmer volume fraction, as well as characterise the distribution of tracers in the suspension.  Finally, we discuss our results and future directions in Section \ref{sec:Discussion}.

\section{Mathematical model for the simulations}
\label{sec:model}
In our simulations, we consider $N_p$ squirmers dispersed in fluid containing $N_t$ tracers giving a total number of particles $N = N_p + N_t$.  All particles are spherical, however the squirmers have radius $a_{sw}$, while the smaller tracers have radius $a$.  The position of particle $n$ is denoted by $\mathbf{Y}_n$, while its orientation is $\mathbf{p}_n$.  The motion of both the spherical tracers and squirmers is governed by overdamped Langevin dynamics, which may be expressed as
\begin{align}
\frac{d\mathcal{Y}}{dt}& = \mathcal{V}_{sq} +\mathcal{V} + \tilde{\mathcal{V}} + k_BT\nabla_{\mathcal{Y}}\cdot \mathcal{M}^{\mathcal{V}\mathcal{F}} \\
\frac{d\mathcal{P}}{dt} &= \mathcal{Q}(\mathcal{W} + \mathcal{W}_{sq} + \tilde{\mathcal{W}}) + k_BT\left(\mathcal{Q} \nabla_{\mathcal{Y}}\cdot \mathcal{M}^{\mathcal{W}\mathcal{F}} - 2 \mathcal{D}^{\mathcal{W}\mathcal{T}} \mathcal{P}\right)
\end{align}
where $\mathcal{Y}$ is the $3N \times 1$ vector containing all particle positions, i.e. $[\mathbf{Y}^T_1, \mathbf{Y}^T_2,\dots, \mathbf{Y}^T_N]^T$, while $\mathcal{P} = [\mathbf{p}^T_1, \mathbf{p}^T_2,\dots, \mathbf{p}^T_N]^T$ is that for the particle orientations.  The matrix $\mathcal{Q}$ is the block diagonal matrix whose non-zero entries are given by $\mathcal{Q}_{3n+k,3n+l} = \epsilon_{klm}\mathbf{p}^n_{m}$ for $n = 1,\dots, N$. 

The vectors $\mathcal{V}$ and $\mathcal{W}$ are the translational and angular velocities, respectively, of the particles that are related to the vector of forces, $\mathcal{F}$, and torques, $\mathcal{T}$, on the particles through
\begin{align}
\left[\begin{array}{c}
\mathcal{V} \\
\mathcal{W} \\
\end{array}\right]=
\left[\begin{array}{cc}
\mathcal{M}^\mathcal{VF} & \mathcal{M}^\mathcal{VT} \\
\mathcal{M}^\mathcal{WF} & \mathcal{M}^\mathcal{WT}  \\
\end{array}\right]
=\mathcal{M}
\left[\begin{array}{c}
\mathcal{F} \\
\mathcal{T} \\
\end{array}\right]
\label{eq:mobrel}
\end{align}
where $\mathcal{M}$ is the $6N \times 6N$ low Reynolds number mobility matrix for all particles.  In Eq. (\ref{eq:mobrel}), we indicate explicitly the four $3N \times 3N$ submatrices relating either forces or torques with translational or angular velocities.    

In addition to $\mathcal{V}$ and $\mathcal{W}$, the particles have velocities $\mathcal{V}_{sq}$ and angular velocities $\mathcal{W}_{sq}$ which encapsulate the swimming velocity, $U \mathbf{p}_n$, of each squirmer, as well as the additional velocities and angular velocities of all particles due to the flows induced by each of the first two squirming modes (\cite{Blake1971,Ishikawa2006}).  For a squirmer centred at the origin and in a frame moving with the squirmer, this induced flow is given by
\begin{align}
\mathbf{u}(\mathbf{x})&= -\frac{B_1}{3}\frac{a^3}{r^3}\left(\mathbf{I} - 3\frac{\mathbf{xx}^T}{r^2}\right)\mathbf{p} + \left(\frac{a^4}{r^4} - \frac{a^2}{r^2}\right)\frac{B_{2}}{2}\left(3 \left(\frac{\mathbf{p}\cdot\mathbf{x}}{r} \right)^2 - 1\right)\frac{\mathbf{x}}{r} \nonumber\\
&- \frac{a^4}{r^4} B_{2}\left(\frac{\mathbf{p}\cdot\mathbf{x}}{r}\right)\left(\mathbf{I} - \frac{\mathbf{xx}^T}{r^2}\right)\mathbf{p}. \label{eq:squ}
\end{align}
where $B_1$ is related to the swimming speed through $B_1 = 3U/2$ (\cite{Blake1971}) and $B_2$ controls the strength and sign of the swimming stresslet, $\mathbf{G}$, through
\begin{align}
\mathbf{G} &= \frac{4}{3}\pi\eta a^2\left(3\mathbf{p}\mathbf{p} - \mathbf{I}\right)B_2 \label{eq:sqstresslet}. 
\end{align}
The squirming parameter $\beta = B_2/B_1$ describes the relative stresslet strength  (\cite{Ishikawa2006}).  For $\beta > 0$, the squirmer is a `puller,' bringing fluid in along $\mathbf{p}$ and expelling it laterally, whereas if $\beta < 0$, the squirmer is a `pusher,' expelling fluid along $\mathbf{p}$ and bringing it in laterally.

The random velocities, $\tilde{\mathcal{V}}$ and angular velocities, $\tilde{\mathcal{W}}$, obey the fluctuation-dissipation theorem and their statistics are related to the mobility matrix $\mathcal{M}$ through
\begin{align}
\langle \tilde{\mathcal{V}}(t)\rangle &= 0\\
\langle \tilde{\mathcal{W}}(t)\rangle &= 0\\
\left\langle \left[\begin{array}{c}
\tilde{\mathcal{V}}(t) \\
\tilde{\mathcal{W}}(t) \\
\end{array}\right] 
\left[\begin{array}{cc}
\tilde{\mathcal{V}}^T(t') & \tilde{\mathcal{W}}^T(t')
\end{array}\right]
\right\rangle &= 2k_BT \mathcal{M}\delta(t-t').
\label{eq:PM2}
\end{align}

Also appearing in the equations of motion are the thermal drift terms $k_BT\nabla_{\mathcal{Y}}\cdot \mathcal{M}^{\mathcal{V}\mathcal{F}}$, $k_BT \mathcal{Q}\nabla_{\mathcal{Y}}\cdot \mathcal{M}^{\mathcal{W}\mathcal{F}}$, and $-2k_BT \mathcal{D}^{\mathcal{W}\mathcal{T}} \mathcal{P}$ that arise from taking the overdamped limit and are required to obtain particle distribution dynamics that are governed by Smoulochowski's equation.  The diagonal matrix $\mathcal{D}^{\mathcal{W}\mathcal{T}}$ has entries $\mathcal{D}^{\mathcal{W}\mathcal{T}}_{ij} = \mathcal{M}^{\mathcal{W}\mathcal{T}}_{ij}$ for $i = j$ and zero otherwise.  
\subsection{Force-coupling method}
To compute the particle dynamics, we rely on the force-coupling method and two recent extensions to active particles (\cite{Delmotte_2015a}) based on the steady squirmer model (\cite{Lighthill1952,Blake1971,Ishikawa2006}) and to Brownian suspensions (\cite{Keaveny2014,Delmotte_2015b}) using concepts from fluctuating hydrodynamics.  The force-coupling method uses regularized force distribution and volume averaging to generate the far-field approximation of the mobility matrix, $\mathcal{M}$.  

To compute the particle velocities and angular velocities arising from the forces and torques on the particles, as well as those due to squirming, we first solve for the Stokes flow
\begin{align}
\bm{\nabla} p - \eta \nabla^2 \mathbf{u} & = \sum_{n=1}^N \mathbf{F}_n \Delta_n(\mathbf{x})  + \sum_{n=1}^{N_p} \mathbf{S}_{n} \cdot \bm{\nabla}\Theta_n(\mathbf{x}) \nonumber\\
&+ \mathbf{G}_{n} \cdot \bm{\nabla} \Delta_n(\mathbf{x}) + \mathbf{H}_n\nabla^2\Theta_n(\mathbf{x}) \label{eq:FCM1}\\
\bm{\nabla}\cdot \mathbf{u}& = 0 \label{eq:FCM2}
\end{align}
where $\mathbf{F}_n$ is the force particle $n$ exerts on the fluid and $\mathbf{S}_{n}$ is the stresslet of particle $n$ due to its rigidity.  We note in our simulations that all particles are torque-free, i.e. $\tau_n = 0$ for all $n$, and we ignore the tracer stresslets due to their small size.  The only forces we consider are short-range pairwise repulsive forces that represent contact forces between two particles, and prevent overlap (\cite{Delmotte_2015a}).  Additionally, for the squirmers, we have 
\begin{align}
\mathbf{H}_n &= -\frac{4}{3}\pi\eta a^3 B_1 \mathbf{p}_n.\\
\mathbf{G}_n &= \frac{4}{3}\pi\eta a^2\left(3\mathbf{p}_n\mathbf{p}_n - \mathbf{I}\right)\beta B_1 \label{eq:sqstresslet} 
\end{align}
as to yield flow corresponding to Eq. \ref{eq:squ} for each squimer.  In Eq. (\ref{eq:FCM1}), we also have the two Gaussian envelopes that are used to project the particle forces onto the fluid,
\begin{eqnarray}
\Delta_n(\mathbf{x})&=&(2\pi\sigma_{n; \Delta}^2)^{-3/2}\textrm{e}^{-|\mathbf{x} - \mathbf{Y}_n|^2/2\sigma_{n;\Delta}^2} \nonumber\\
\Theta_n(\mathbf{x})&=&(2\pi\sigma_{n;\Theta}^2)^{-3/2}\textrm{e}^{-|\mathbf{x} - \mathbf{Y}_n|^2/2\sigma_{n;\Theta}^2}.
\label{eq:FCM2}
\end{eqnarray} 
The length scales $\sigma_{n;\Delta}$ and $\sigma_{n;\Theta}$ are related to the radius, $a_n$, of particle $n$ through $\sigma_{n;\Delta} = a_n/\sqrt{\pi}$ and $\sigma_{n;\Theta} = a_n/\left(6\sqrt{\pi}\right)^{1/3}$.  After solving Eq. (\ref{eq:FCM1}) for $\mathbf{u}$, the velocity of each tracer is determined from 
\begin{align}
\mathbf{V}_n&=\int\mathbf{u}\Delta_n(\mathbf{x})d^3\mathbf{x} \label{eq:FCM3a},
\end{align}
while for the squirmers, the velocities, angular velocities, and local-rates-of-strain are given by
\begin{align}
\mathbf{V}_n &= U\mathbf{p}_n - \mathbf{W}_n + \int \mathbf{u} \Delta_n(\mathbf{x}) d^3\mathbf{x} \label{eq:part_vel_self_ind}\\
\bm{\Omega}_n &= \frac{1}{2}\int\left[\bm{\nabla}\times\mathbf{u}\right] \Theta_n(\mathbf{x})d^3\mathbf{x} \label{eq:part_rot}\\
\mathbf{E}_n&=-\mathbf{K}_n + \frac{1}{2}\int \left[\bm{\nabla}\mathbf{u} + (\bm{\nabla}\mathbf{u})^T\right]\Theta_n(\mathbf{x})d^3\mathbf{x}. \label{eq:ROS_self_ind} = 0
\end{align}
where the terms $\mathbf{W}_n$ and $\mathbf{K}_n$ are included to subtract away artificial self-induced velocities and local-rates-of-strain due to the volume integration of the squirming modes.  Expressions for these terms are provided in \cite{Delmotte_2015a}.  The local rate-of-strain for each squirmer is required to be zero, $\mathbf{E}_n = 0$, giving rise to the stresslets, $\mathbf{S}_n$.

To compute the random velocities and angular velocities, we consider the Stokes flow
\begin{align}
\bm{\nabla} p - \eta \nabla^2 \tilde{\mathbf{u}} &= \bm{\nabla}\cdot \mathbf{P} + \sum_{n=1}^{N_p} \tilde{\mathbf{S}}_{n} \cdot \bm{\nabla}\Theta_n(\mathbf{x})  \nonumber\\
 \bm{\nabla}\cdot \tilde{\mathbf{u}} &= 0.
\label{eq:RanVel1}
\end{align}
where $\mathbf{P}$ is a fluctuating stress driving the random fluid flow.  The statistics for $\mathbf{P}$, in index notation, are given by
\begin{align}
	\left\langle P_{jl}\right\rangle&=0\\
	\left\langle P_{jl}(\mathbf{x},t)P_{pq}(\mathbf{x}',t') \right\rangle&=2k_BT\eta\left(\delta_{jp}\delta_{lq} + \delta_{jq}\delta_{lp}\right)\delta(\mathbf{x}-\mathbf{x}')\delta(t-t'). 
\label{eq:RanVel2}
\end{align}
Using the FCM volume averaging operators, we first enforce 
\begin{align}
\tilde{\mathbf{E}}_n&= \frac{1}{2}\int \left[\bm{\nabla}\tilde{\mathbf{u}} + (\bm{\nabla}\tilde{\mathbf{u}})^T\right]\Theta_n(\mathbf{x})d^3\mathbf{x} = 0
\label{eq:ROS_self_ind}
\end{align}
to obtain the values of $\tilde{\mathbf{S}}_n$ and subsequently compute
\begin{align}
\tilde{\mathbf{V}}_n &= \int \tilde{\mathbf{u}} \Delta_n(\mathbf{x}) d^3\mathbf{x} \label{eq:randvel}\\
\tilde{\bm{\Omega}}_n &= \frac{1}{2}\int\left[\bm{\nabla}\times\tilde{\mathbf{u}}\right] \Theta_n(\mathbf{x})d^3\mathbf{x} \label{eq:randrot}
\end{align}
As demonstrated in \cite{Keaveny2014}, the resulting velocities and angular velocities will satisfy the fluctuation-dissipation theorem with the covariance given by the FCM approximation of the mobility matrix.  We note that while we have described the computation of deterministic and stochastic velocities separately, due to the linearity of Stokes equations, these can be combined into a single computation of the total particle velocities.

While this describes how we obtain the deterministic and random motions of the particles, it remains to integrate the equations of motion with the correct thermal drift.  We accomplish this by employing the midpoint drifter-corrector (DC) time integration scheme (\cite{Delmotte_2015b}) that inherently accounts for the drift terms without having to compute them directly.  In words, the DC recovers the drift by initially advancing the particle positions to the midstep using only the random velocities determined from the fluctuating fluid flow with no stresslets for particle rigidity.  At the midstep, the full computation is performed to find the deterministic velocities, as well as new random particle velocities using the same initial realisation of the fluctuating stress.  The result is correlated motion between the initial and midstep velocities that then can be exploited to capture the correct thermal drift.  For details of the scheme, including error expansions showing explicitly that the drift is recovered, the reader is referred to \cite{Delmotte_2015b}.  

\subsection{Simulations parameters and set-up}
\begin{figure}
\centering
\includegraphics[width=0.95\columnwidth]{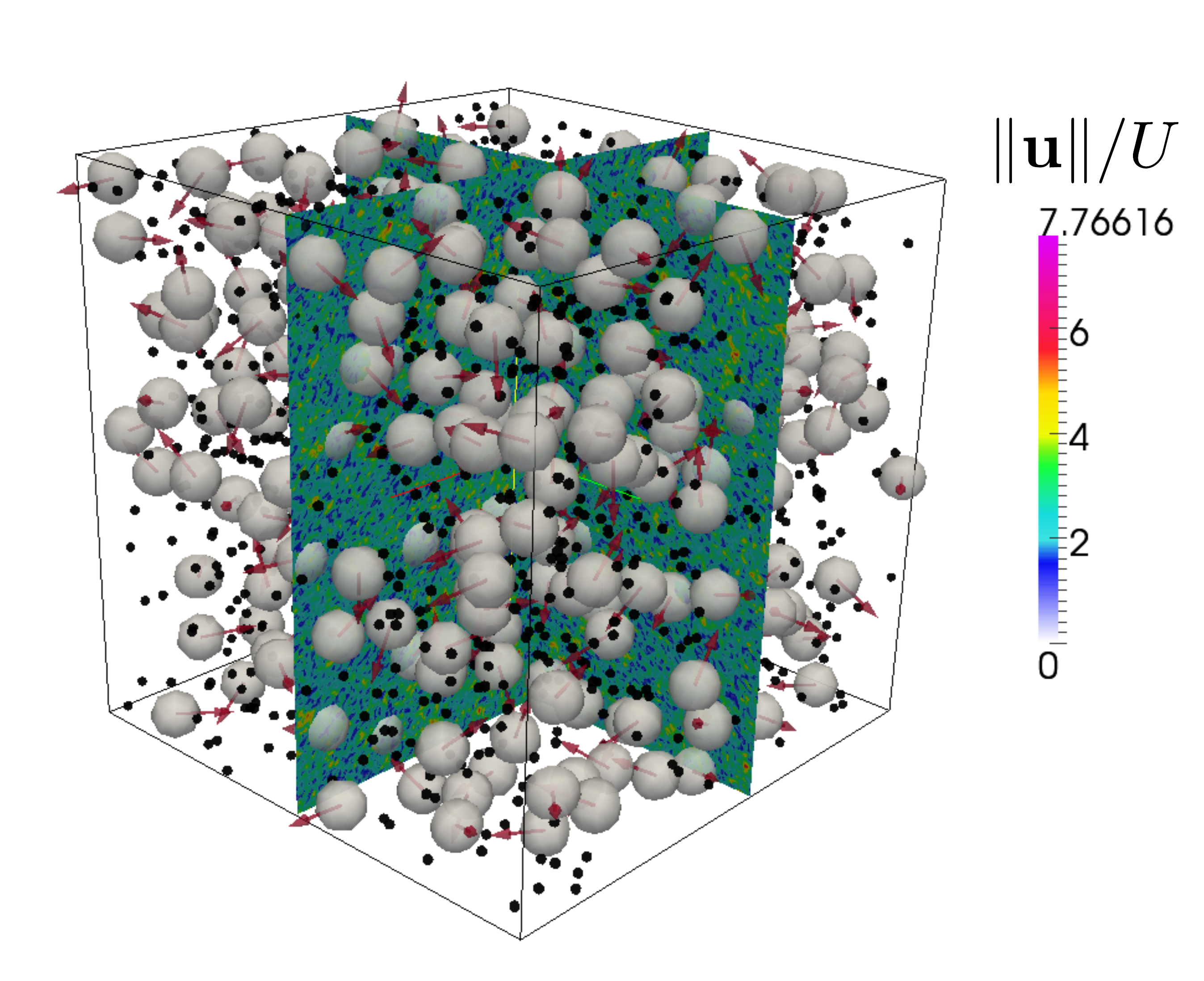}
\caption{\label{fig:snapshot_concentrated_tracers} Snapshot of the simulation domain containing $N_p=301$ swimmers and $1255$ tracers, with $a_{sw}/a =5$. The volume fraction is $\phi_v = 10\%$. The large grey spheres are the squirmers and the small black dots correspond to  the tracers. Vectors represent the swimmers' orientations $\mathbf{p}^n, \, n=1,..,N_p$. Slices show the norm of the fluid velocity field normalized by the intrinsic swimming speed $\|\mathbf{u}\|/U$. One can observe the fluid velocity fluctuations arising from the fluctuating stress.
}
\end{figure}

In our simulations, we set our parameters to match the \emph{C. rheindardtii} experiments of \cite{Leptos2009}.  We set  the swimmer to tracer radius ratio to $a_{sw}/a = 5$ and set the swimming speed, and hence $B_1$, to match $U = 100\mu m s^{-1} = 20 a_{sw}s^{-1}$.  Following \cite{Thiffeault2014}, we set $\beta = 0.5$.  The simulations are carried out in a triply periodic domain with edge length $L = 23a_{sw}$.  The Stokes equations are solved using a Fourier spectral method with $N_{g} = 192$ grid points in each direction.   The number of tracers is set to $N_t = 1255$, resulting in a tracer volume fraction of less than 0.34\%.  The number of swimmers is varied from $N_{p} = 12$ to $451$ to examine very dilute swimmer suspensions of $\phi_{v} = 0.4\%$, as well as semi-dilute cases where $\phi_{v} = 15\%$.  Finally, the thermal energy $k_BT$ is set to obtain the same distribution of tracers displacements as \cite{Leptos2009} in the absence of swimmers.  We note that the resulting value of the diffusion, $D_0 = 0.34 \mu^2s^{-1}$ is slightly higher than the value $D_0 = 0.28 \mu^2s^{-1}$ given in \cite{Leptos2009}.   A snapshot taken from a representative simulation is shown in Figure \ref{fig:snapshot_concentrated_tracers}.

\section{Results}

\subsection{Dilute regime}
\label{sec:Dilute}
\subsubsection{Tracer displacements}
In their experiments, \cite{Leptos2009} measured the time-dependent probability distribution function, $P(\Delta x, \Delta t)$, for tracer displacements for suspensions with swimmer volume fractions $\phi_v = 0-2.2\%$ over $0.3$s.  As the beat period for the flagella of \emph{C. rheindardtii} is $T = 0.02$s, this observation time corresponds to 15 beat cycles.  They find that unlike previous experiments using bacterial baths (\cite{Wu2000}), the tracer displacement distributions exhibit non-Gaussian tails.  The non-Gaussian tails are attributed to rare entrainment events that occur when a tracer particle comes in close proximity to a swimmer's surface.  
\begin{figure}
\begin{centering}
\subfloat[PDF for tracer displacements at time $\Delta t = 0.12s$ for swimmer volume fractions $\phi_v = 0 - 2.2\%$. Symbols represent the data from \cite{Leptos2009}. Solid lines correspond to the simulations.]{\label{fig:PDF_disp_beta_0_5} \includegraphics[width=0.5\columnwidth]{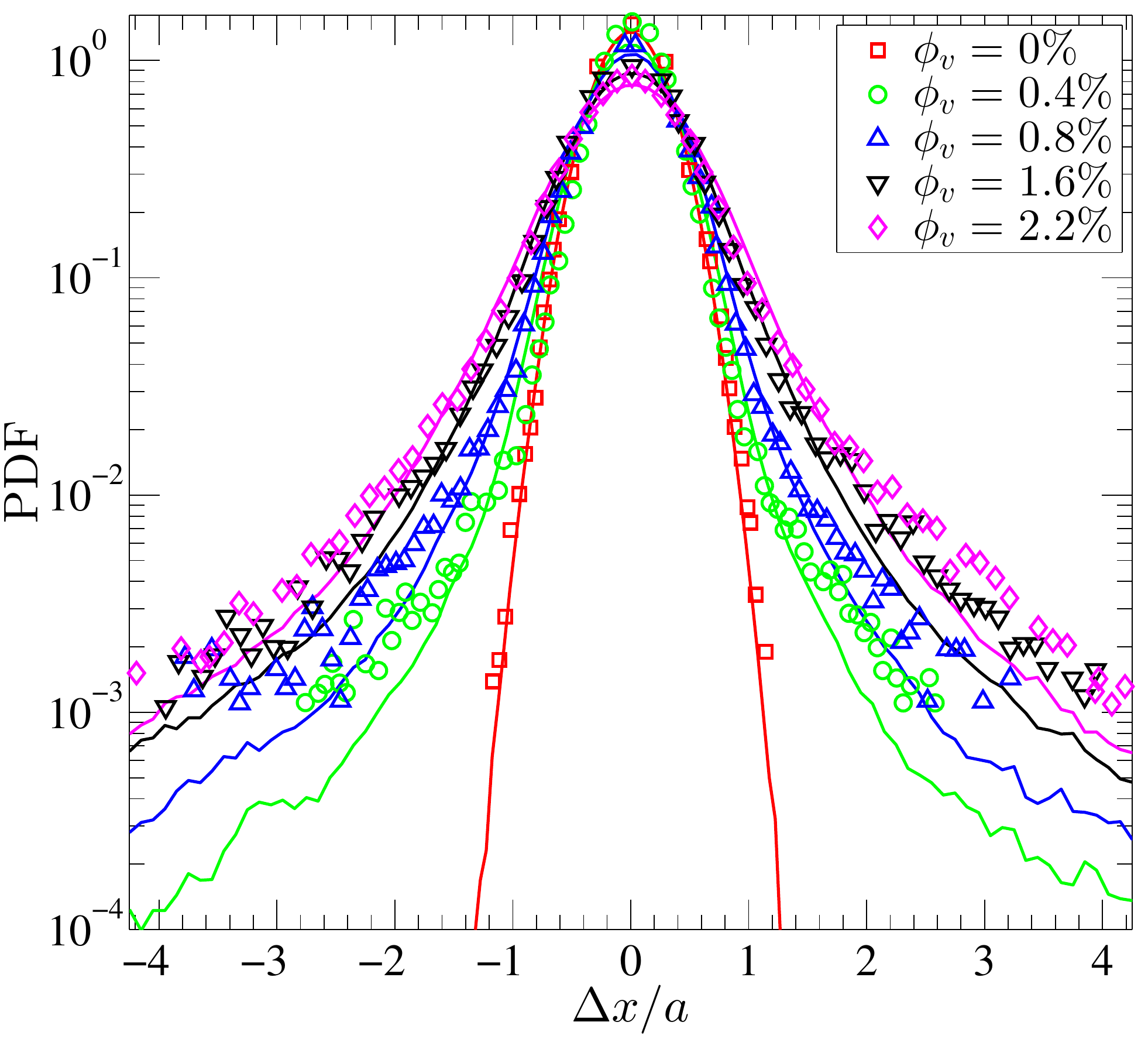}}
\hspace{0.5cm}
\subfloat[Diffusive scaling of the tracer displacement PDF for $\phi_v = 2.2\%$ from the simulations]{\label{fig:PDF_disp_beta_0_5_diff_scaling} \includegraphics[width=0.5\columnwidth]{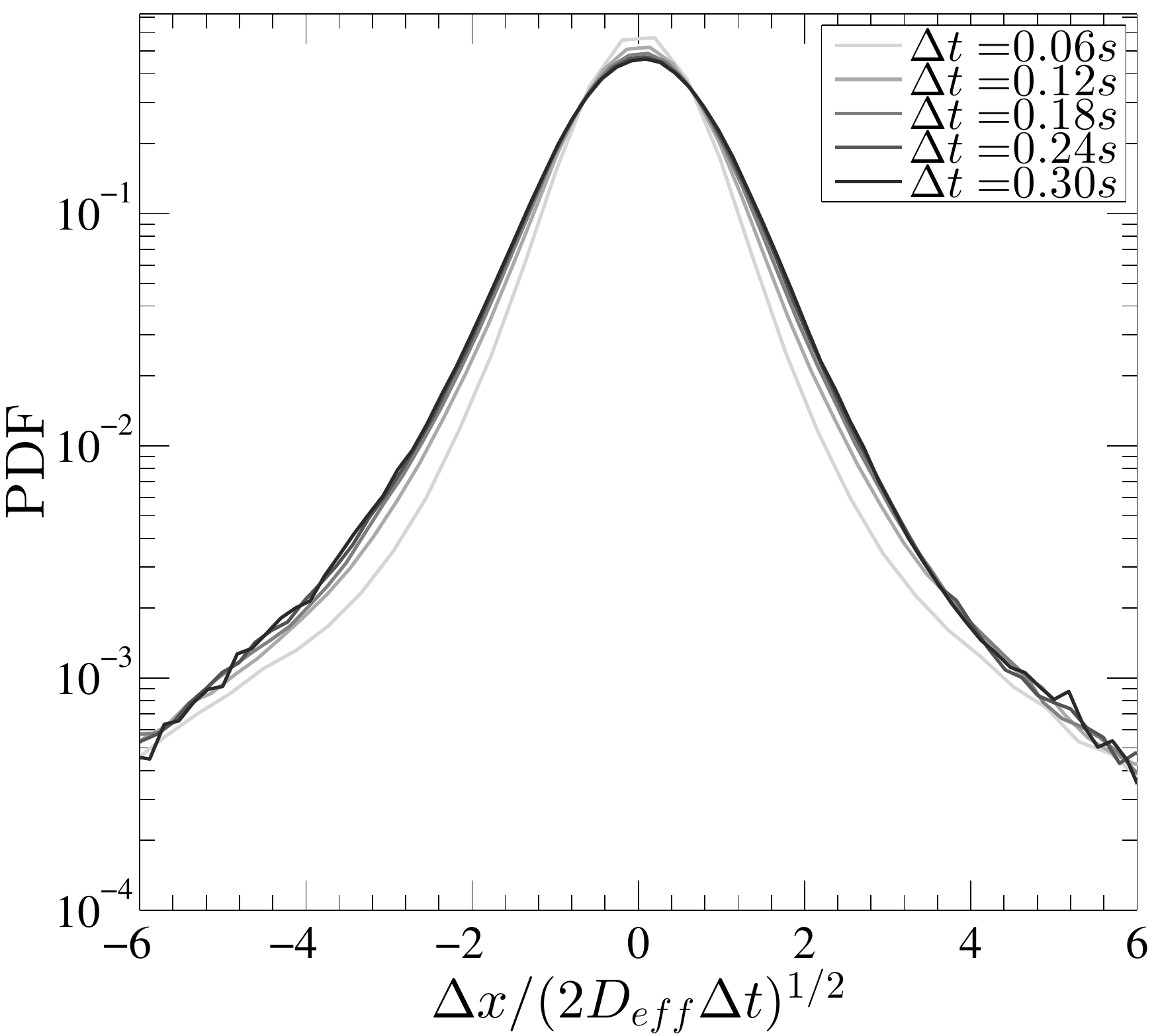}}
\end{centering}
\caption{Tracer displacements in dilute suspensions of squirmers with $\beta = 0.5$. Displacements are averaged over the three spatial directions.
}
\end{figure}
We performed simulations corresponding to the same swimmer volume fractions and observation times as in the experiments of  \cite{Leptos2009}.  The resulting tracer displacement distributions from our simulations are shown in Fig. \ref{fig:PDF_disp_beta_0_5}.  Our squirmer simulations  	adequately   capture the Gaussian core of the distributions.  As in \cite{Leptos2009}, we find that the distributions are self-similar with respect to the diffusive scaling $\Delta x/(2D_{eff}\Delta t)^{1/2}$ (Fig. \ref{fig:PDF_disp_beta_0_5_diff_scaling}).  We find, however, that the larger, but rarer, displacement events related to the tails are slightly underestimated by the model.  This is consistent with similar findings of tracer displacements by squirmers (\cite{Thiffeault2014}).  We attribute this to the fact that in the near-field, the squirmer does not replicate the flow induced by swimming \emph{C. Rheinardtii} and the tails of the PDF for short-time tracer displacements depend on the details of the flow near the swimming micro-organism.  The differences in entrainment due to differences in swimming behaviour have been examined previously in \cite{Pushkin2013}. 

\begin{figure}
\centering
\includegraphics[width=0.95\columnwidth]{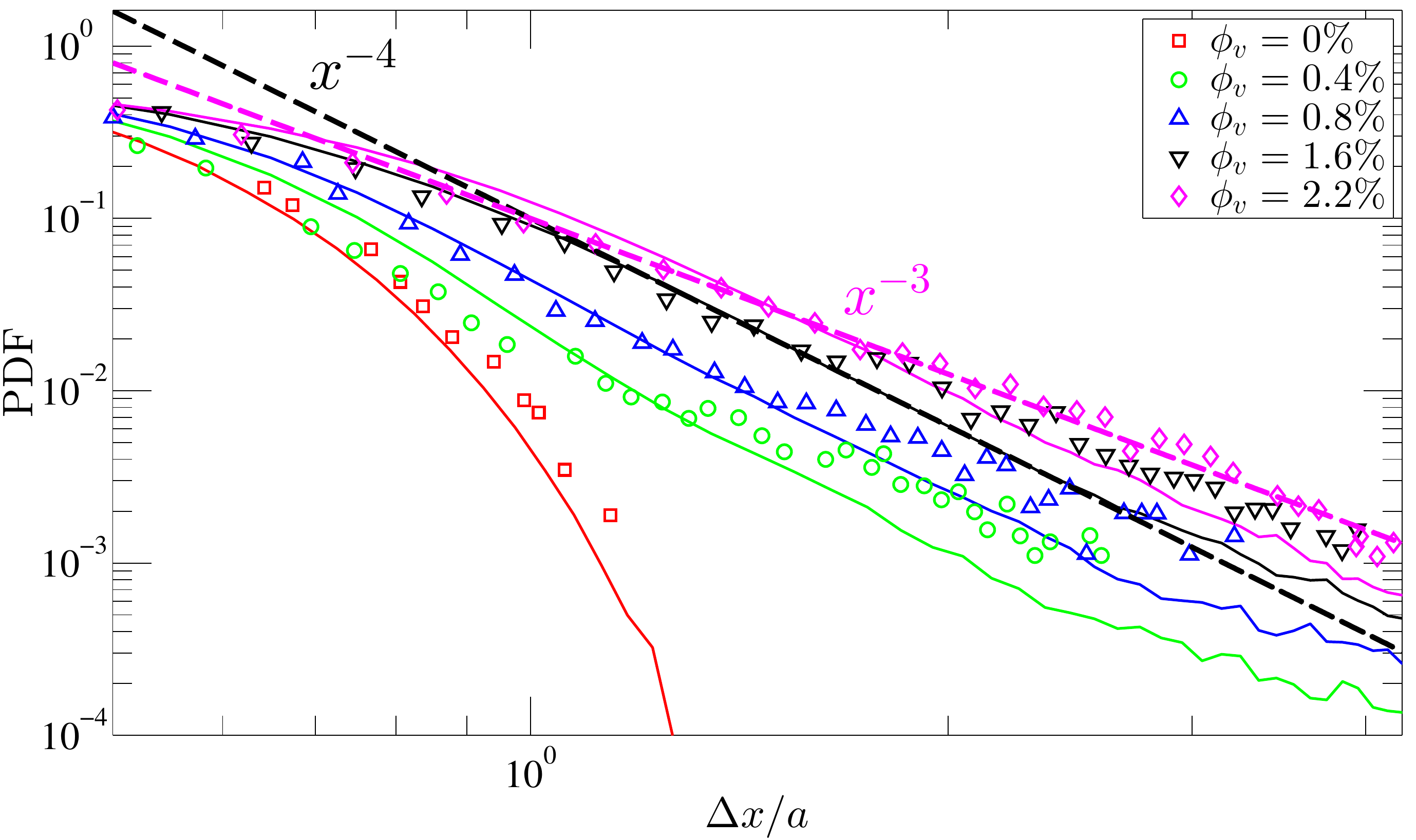}
\caption{\label{fig:Tails_pdf_log_log} The PDF of tracer displacements in log-scale. Symbols represent the data from \cite{Leptos2009}. Solid lines correspond to the simulations.  The dashed lines correspond to the power laws: 
\symbol{\dashed}{}{10}{0}{black}{black}: $x^{-4}$. 
\symbol{\dashed}{}{10}{0}{magenta}{black}: $\textcolor{magenta}{x^{-3}}$.
}
\end{figure}

Fig. \ref{fig:Tails_pdf_log_log} shows the distribution tails in more detail.  Though over less than a decade of displacements, we observe that our simulation data follows a $x^{-4}$ power-law, while the data from \cite{Leptos2009} appears to behave as $x^{-3}$.  We note that \cite{Leptos2009} originally fit the non-Gaussian tails by exponentials.  The $x^{-4}$ decay was observed in experiments by \cite{Kurtuldu2011} and predicted by \cite{Pushkin2014} for suspensions of dipolar swimmmers with random orientations.  

\begin{figure}
\centering
\subfloat[PDF for tracer displacements at times $\Delta t = 0.06-2s$ for $\phi_v = 2.2\%$.]{\label{fig:PDF_disp_beta_0_5_phi_2_2_long_time} \includegraphics[width=0.45\columnwidth]{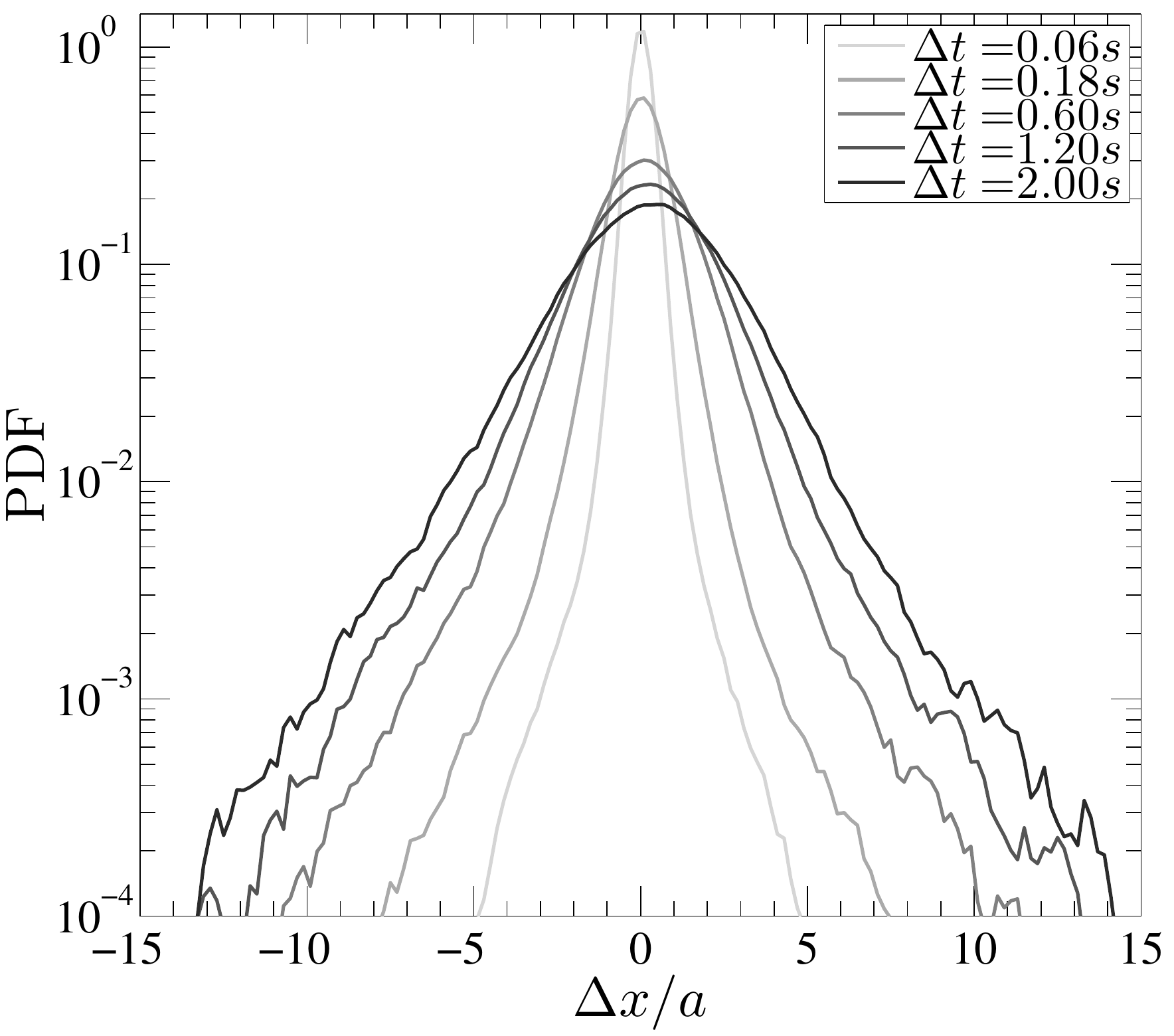}}
\hspace{0.5cm}
\subfloat[Diffusive scaling of the PDF for tracer displacements for $\phi_v = 2.2\%$.]{\label{fig:PDF_disp_beta_0_5_diff_scaling_phi_2_2_long_time} \includegraphics[width=0.45\columnwidth]{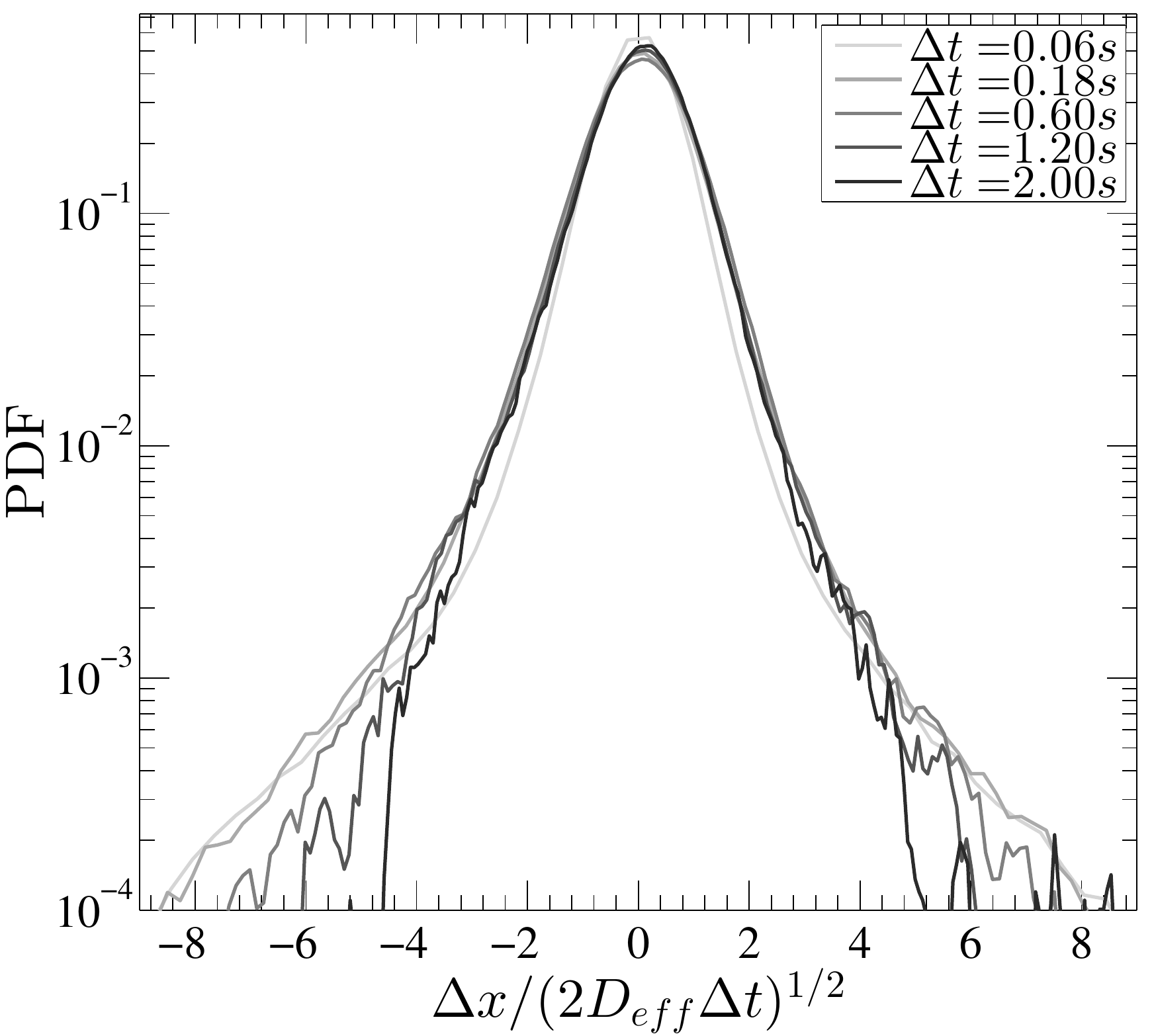}}\\
\subfloat[PDF for tracer displacements at time $\Delta t = 2s$ for various swimmer volume fractions $\phi_v = 0 - 2.2\%$.]{\label{fig:PDF_disp_beta_0_5_long_time} \includegraphics[width=0.5\columnwidth]{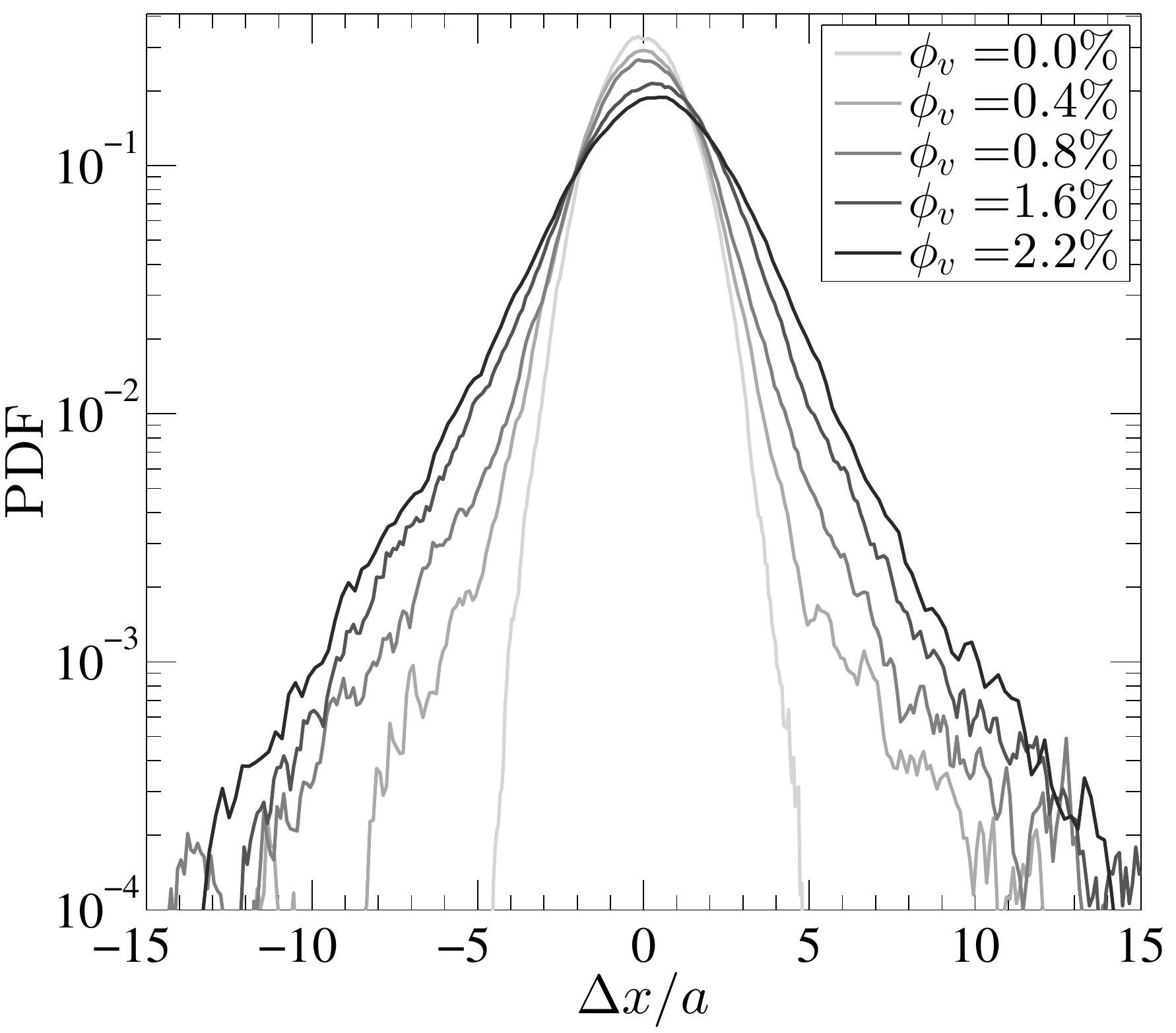}}
\caption{\label{fig:Long_time_dilute_tracers} PDF of tracer displacements for long times.  Displacements are averaged over the three spatial directions.
}
\end{figure}

Previous studies (\cite{Zaid2011,Thiffeault2014}) have argued that the non-Gaussian tails are due to the shortness of the observation time coupled with the rarity of the entrainment events.  To verify this assertion, we have computed the evolution of the tracer displacement PDF over a longer time $\Delta t = 0 - 2s$ as shown in Figure \ref{fig:PDF_disp_beta_0_5_phi_2_2_long_time} for $\phi_v = 2.2\%$.  As $\Delta t$ increases, we see that the Gaussian core of the distribution broadens.  We also observe that small shifts in the mean value appears, which we attribute to statistical fluctuations in the data.  After rescaling by the diffusive timescale, we see in Figure \ref{fig:PDF_disp_beta_0_5_diff_scaling_phi_2_2_long_time} that the tails decay more rapidly and the distribution approaches a Gaussian as $\Delta t$ increases.  This convergence to a Gaussian is observed for any volume fraction $\phi_v = 0-2.2\%$, see Figure \ref{fig:PDF_disp_beta_0_5_long_time}, with the rate of convergence increasing as $\phi_v$ increases.  \cite{Thiffeault2014} derived the criterion for reaching a Gaussian distribution. He showed that for spherical squirmers with $\beta = 0.5$, the time to reach Gaussianity is $\Delta t = 3.57, 1.74, 0.8, 0.5 s$ for $\phi_v = 0.4, 0.8, 1.6, 2.2\%$ respectively.  Our simulation data (Figure \ref{fig:Long_time_dilute_tracers}) are in inaccordance with these results.  We also note that convergence to a Gaussian distribution in dilute suspensions was observed in experiments  where the algal suspension was confined to a liquid film (\cite{Kurtuldu2011}).

While our results point towards convergence to a Gaussian distribution, we note that in \cite{Zaid2011}, using point sources with no steric interactions, they found that Gaussianity should arise only when the disturbance flow induced by the swimmers decays as $r^{-n}$ with $n=1$, and if $n\geq 2$, as is the case of squirmers, the fluid velocity distribution, and thus tracer displacements, should deviate from Gaussianity, even at long times.  Their results are supported by the work of \cite{Rushkin2010} on dilute suspensions of \emph{Volvox} ($\phi_v \leq 1.5\%$). Their study shows that when considering only settling forces ($n=1$), the fluid velocity fluctuations follow a normal distribution.  When accounting for the degenerate quadrupole due to ciliary beating ($n=3$), the fluid velocity fluctuations exhibit strong deviations from Gaussianity, as confirmed by experimental data.

\subsubsection{Enhanced Diffusion}

\begin{figure}
\begin{centering}
\subfloat[Mean-squared displacement over time.]{ \includegraphics[width=0.5\columnwidth]{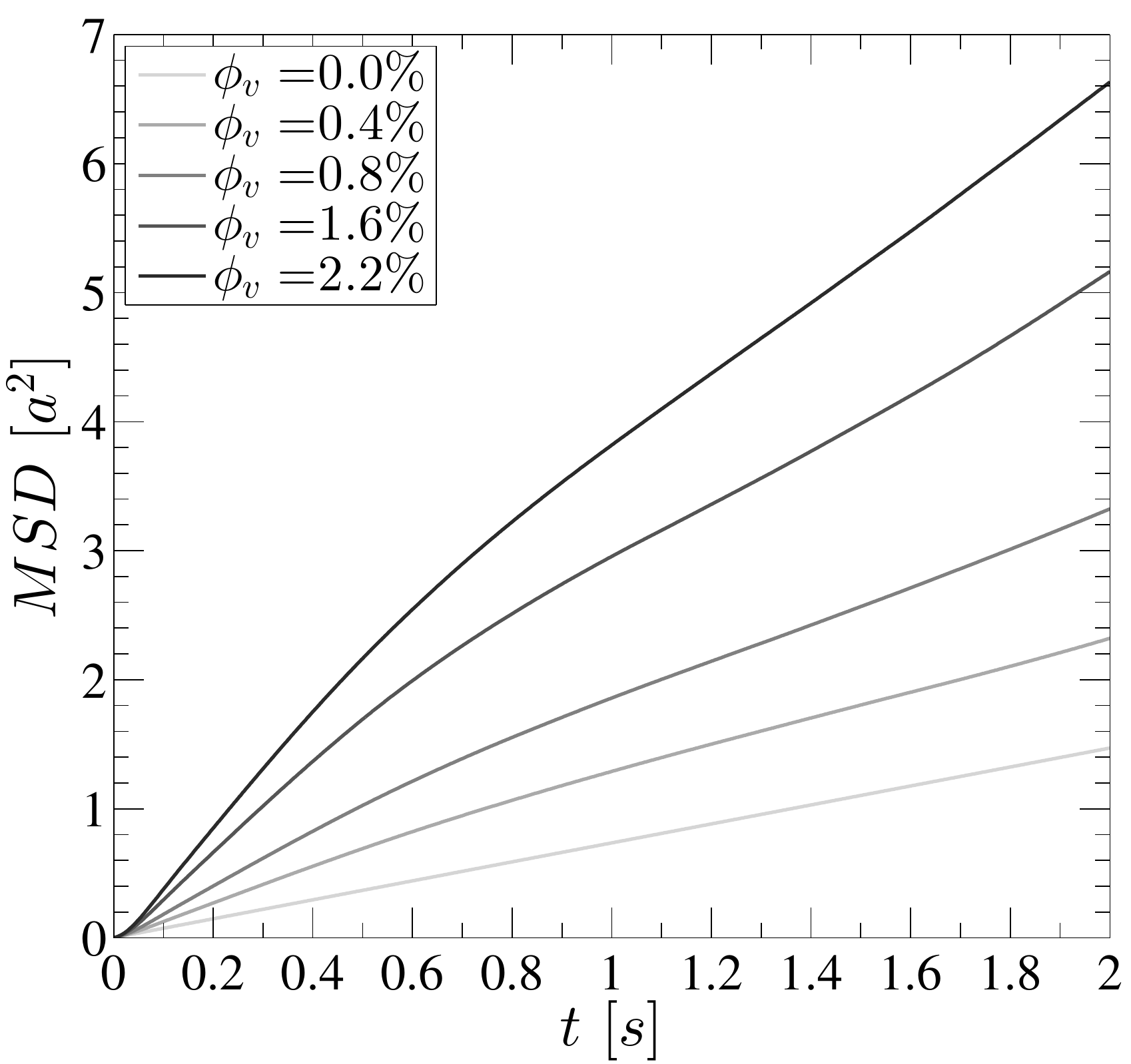}}
\subfloat[Mean-squared displacement over time in log scale.]{ \includegraphics[width=0.5\columnwidth]{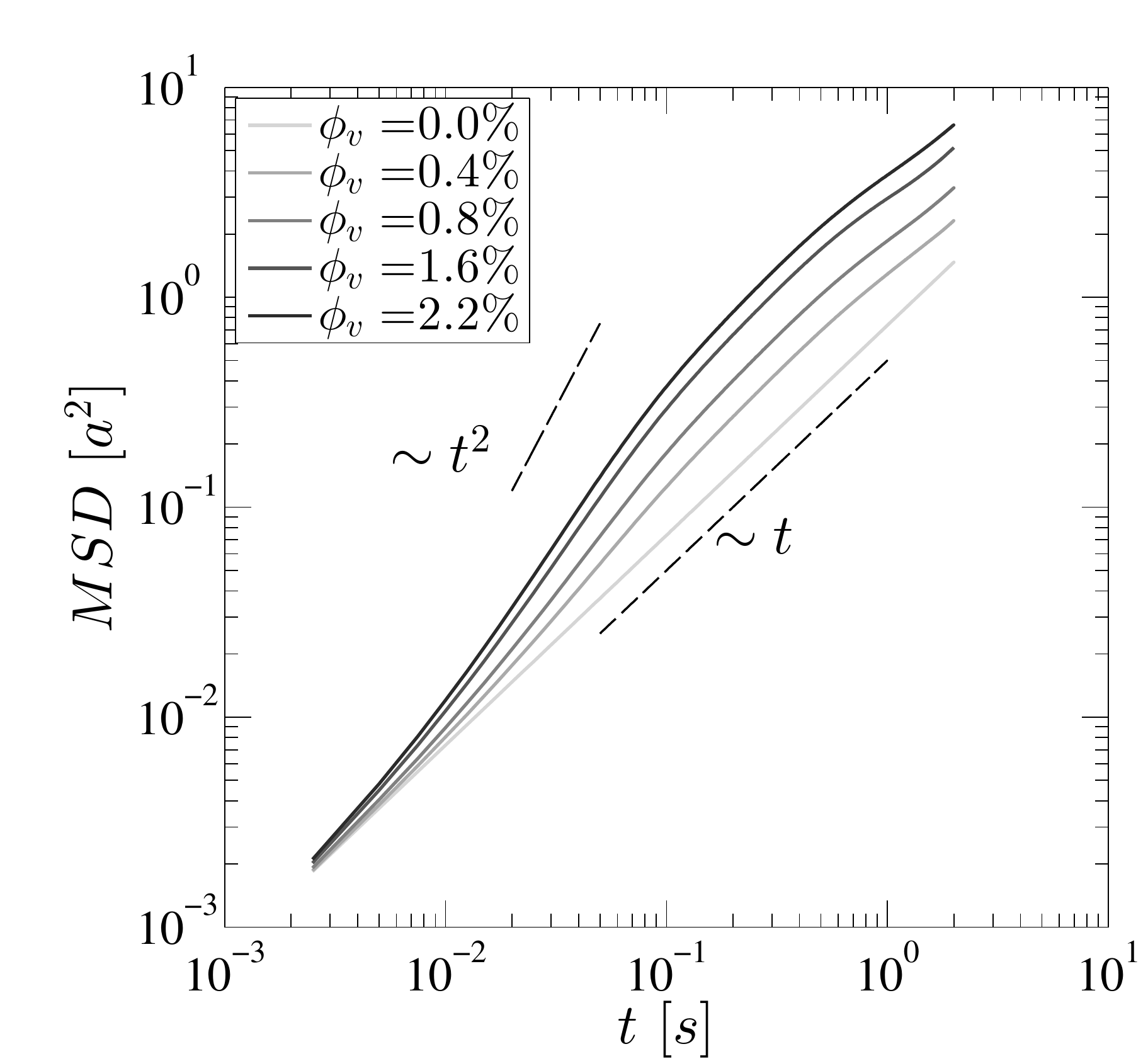}}
\end{centering}
\caption{\label{fig:MSD_beta_0_5} Mean-squared displacement of tracers for different swimmer concentrations.  Displacements are averaged over the three spatial directions.
}
\end{figure}

\begin{figure}
\begin{centering}
\subfloat{\includegraphics[width=0.5\columnwidth]{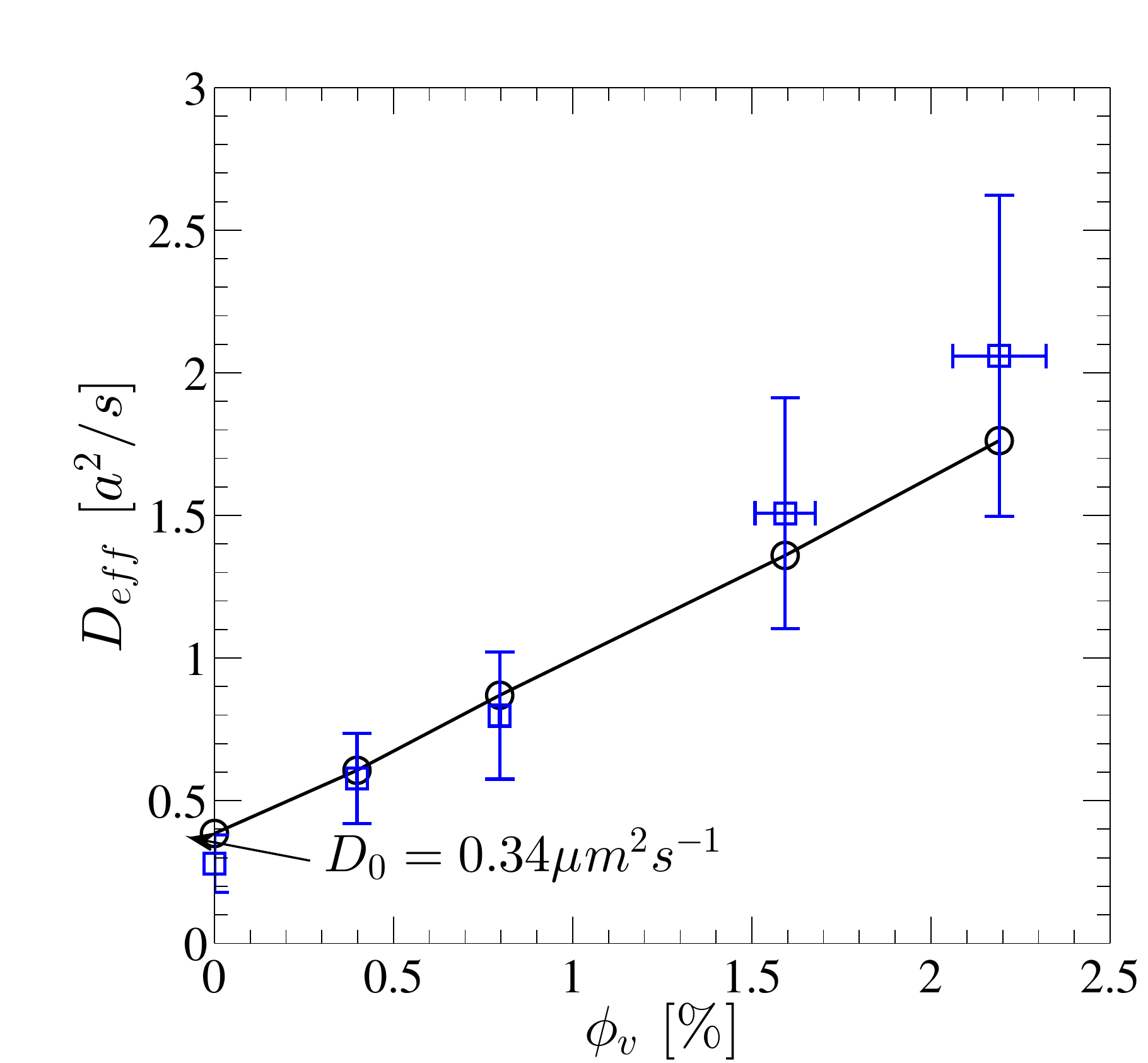}}
\caption{\label{fig:Deff_beta_0_5_kbts} Effective diffusion coefficient of tracers for the squirmer model with $\beta = 0.5$. 
\symbol{\solid}{\bigcircle}{22}{0}{black}{black}: simulations. 
\symbol{}{\ssquareb}{0}{0}{black}{blue}: data from \cite{Leptos2009}.
}
\end{centering}
\end{figure}

In addition to the tracer displacement distribution, \cite{Leptos2009} measured the mean-squared displacement of tracers as a function of time and observed that the motion of tracers is diffusive obeying $\langle\Delta x ^2\rangle = 2 D_{eff} t$ for all times.  Figure \ref{fig:MSD_beta_0_5} shows the mean-squared tracer displacement from our squirmer simulations.  As $\phi_v$ increases, we observe that the onset of the diffusive regime occurs after a period of anomalous transport at short times where $\langle\Delta x ^2\rangle \sim t^{\Theta}, \, 1<\Theta<2$.  We note that similar behaviour was observed in other experiments  (\cite{Wu2000,Kurtuldu2011,Kurihara_2017}) and the theoretical prediction ${\Theta}=3/2$ has also been proposed (\cite{Kurihara_2017}).  Once we have reached the diffusive regime at $\Delta t = 2s$, however, our values for the mean-squared displacement are very close to those given in \cite{Leptos2009}.
We extract the effective diffusion coefficient from our data by fitting the mean-squared displacement by the solution of the Langevin equation (\cite{Wu2000})
\begin{equation}
 \langle \Delta x ^2 \rangle = 2D_{eff}\Delta t\left[1-\exp(-t/\tau) \right]
\end{equation}
where $\tau$ is the timescale over which ballistic motion ($\langle \Delta x ^2 \rangle \sim2 \frac{D_{eff}}{\tau}t^2$) transitions to diffusive behavior $(\langle\Delta x ^2\rangle \sim 2 D_{eff} t)$.  In Figure \ref{fig:Deff_beta_0_5_kbts}, we compare the effective diffusion coefficient from our simulations with those obtained by \cite{Leptos2009}.  We can see that the simulated values match the experimental ones within the statistical errors reported for the experiments.

\subsubsection{Steric vs. hydrodynamic interactions}

In our simulations, and in all experiments, the tracer particles have a finite size and experience contact forces with nearby swimmers.  Hence, three phenomena contribute to the tracer diffusivity in the dilute regime: the flows generated by the swimmers, collisions with swimmers due to contact forces, and tracer Brownian motion.  In order to quantify the effect of each, we consider three situations: (i) a bidisperse suspension of passive particles achieved by setting $U = 0, \mathbf{H}_n = \mathbf{0},$ and $\mathbf{G}_n = \mathbf{0}$ for all swimmers,  (ii) a suspension where the swimmers move in the fluid without generating any swimming disturbances and interacting with the tracers through contact forces ($U \neq 0, \mathbf{H}_n = \mathbf{0},$ $\mathbf{G}_n = \mathbf{0}$), and (iii) the full simulation model.

\begin{figure}
\begin{centering}
 \includegraphics[width=0.5\columnwidth]{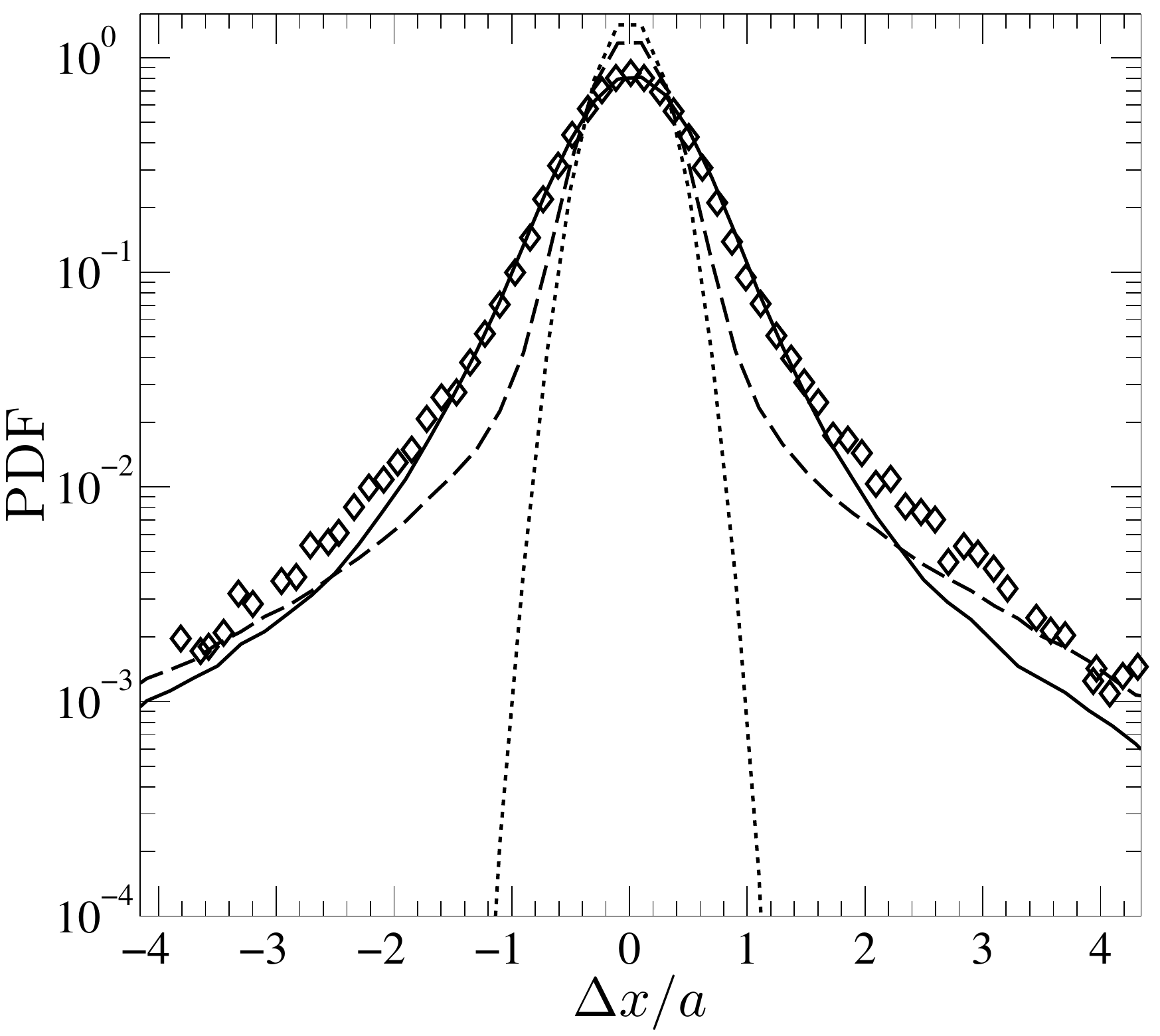}
\caption{\label{fig:steric_vs_hydro} PDF for tracer displacements for $\Delta t = 0.12s$ for swimmer volume fraction $\phi_v = 2.2 \%$ and $\beta = 0.5$. Dotted line: case (i), swimmers replaced by passive particles, Dashed line: case (iI), moving swimmers that do not generate flow, Solid line: case (iii), full model, Symbols: experiments from \cite{Leptos2009}.}
\end{centering}
\end{figure}

Figure \ref{fig:steric_vs_hydro} shows the resulting PDF of tracer displacements for a dilute suspension ($\phi_v = 2.2 \%$) of squirmers with $\beta = 0.5$. 
In the purely passive case (i), the PDF is Gaussian as expected.  For case (ii), the motion of the larger particles and collisions with the tracers generate a non-Gaussian PDF.  The Gaussian core of this PDF is similar to that observed in case (i), but we see that there are additional non-Gaussian tails with power-law decay.  When moving to the full model (case (iii)), the Gaussian core widens and the power-law decay of the tails increases.  This widening of the Gaussian core was also seen in the experiments.  These results show that rare swimmer-tracer collisions produce the large displacements that lead to the non-Gaussian tails, while the flows generated by the swimmers act to broaden the Gaussian core, thus enhancing the diffusivity of the tracer particles at short times.

\begin{figure}
\centering
\includegraphics[width=0.5\columnwidth]{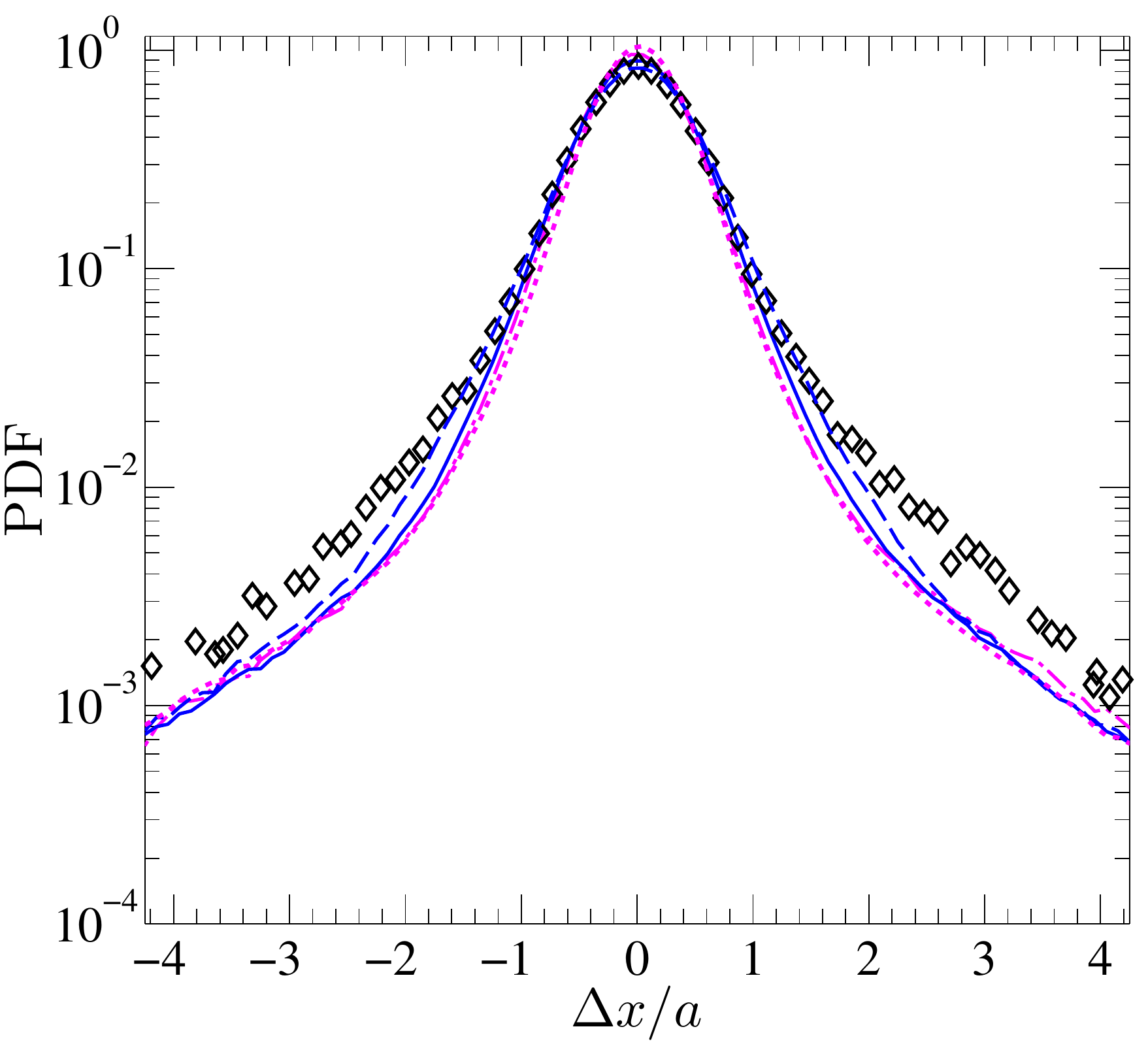}
\caption{\label{fig:Tracers_disp_time_dep} PDF of tracer displacements at $\Delta t = 0.12s$ for the time-dependent and steady models with $\phi_v = 2.2\%$. 
\symbol{}{\losange}{0}{0}{black}{black}: data from \cite{Leptos2009}. 
\symbol{\solid}{}{0}{0}{blue}{black}: $\bar{\beta} = 0.5$, time-dependent. 
\symbol{\dashed}{}{0}{0}{blue}{black}: $\beta = 0.5$, steady. 
\symbol{\dashdot}{}{0}{0}{magenta}{black}: $\bar{\beta} = 0.12$, time-dependent. 
\color{magenta}{\tdot \tdot \tdot \tdot \tdot \tdot} \color{black}{:  $\beta = 0.12$, steady.} 
}
\end{figure}

\subsubsection{Role of time-dependence}
The flows generated by \emph{C. rheinardtii} are time-dependent (\cite{Guasto2010}) due to the beating of its two flagella.  We can include this time dependence by allowing the parameters $B_1$ and $B_2$ appearing in the steady squimer model to be periodic functions of time.  These functions can then be tuned to match the time-dependent swimming speed of \emph{C. rheinardtii} and the location of the flow stagnation point as measured in \cite{Guasto2010}.  The details of our tuning procedure can be found in \cite{Delmotte_2015a}. After tuning, we find the average value of the squirming parameter over one beat period is $\bar{\beta} = 0.12$.  

Figure \ref{fig:Tracers_disp_time_dep} shows the PDF of tracer displacements for $\phi_v = 2.2$\% from the fully time-dependent model.  For comparison, we have also included results from simulations of the time-dependent model with $\bar{\beta} = 0.5$, as well as results from the steady model with $\beta = 0.12$ and $\beta=0.5$.  For $\beta = 0.12$, we find that the steady and time-dependent PDFs are nearly indistinguishable.  As $\bar{\beta}$ increases to $\bar{\beta} = 0.5$, the PDF is slightly narrower with respect to that of its corresponding steady simulation $\beta = 0.5$.  The difference, however, is quite small, and we find that time-dependence of the squirming modes does not significantly affect the PDF of the tracer displacements. These results are in accordance with \cite{Dunkel2010} that showed in the dilute regime the time-dependent and stroke-averaged swimming disturbances provide similar tracer scattering.

\subsection{Semi-dilute regime}
\label{sec:tracer_concentrated}



\begin{figure}
\begin{centering}
\subfloat[$\Delta t = 0.3s$]{\label{fig:PDF_disp_concentrated_DT_03} \includegraphics[width=0.5\columnwidth]{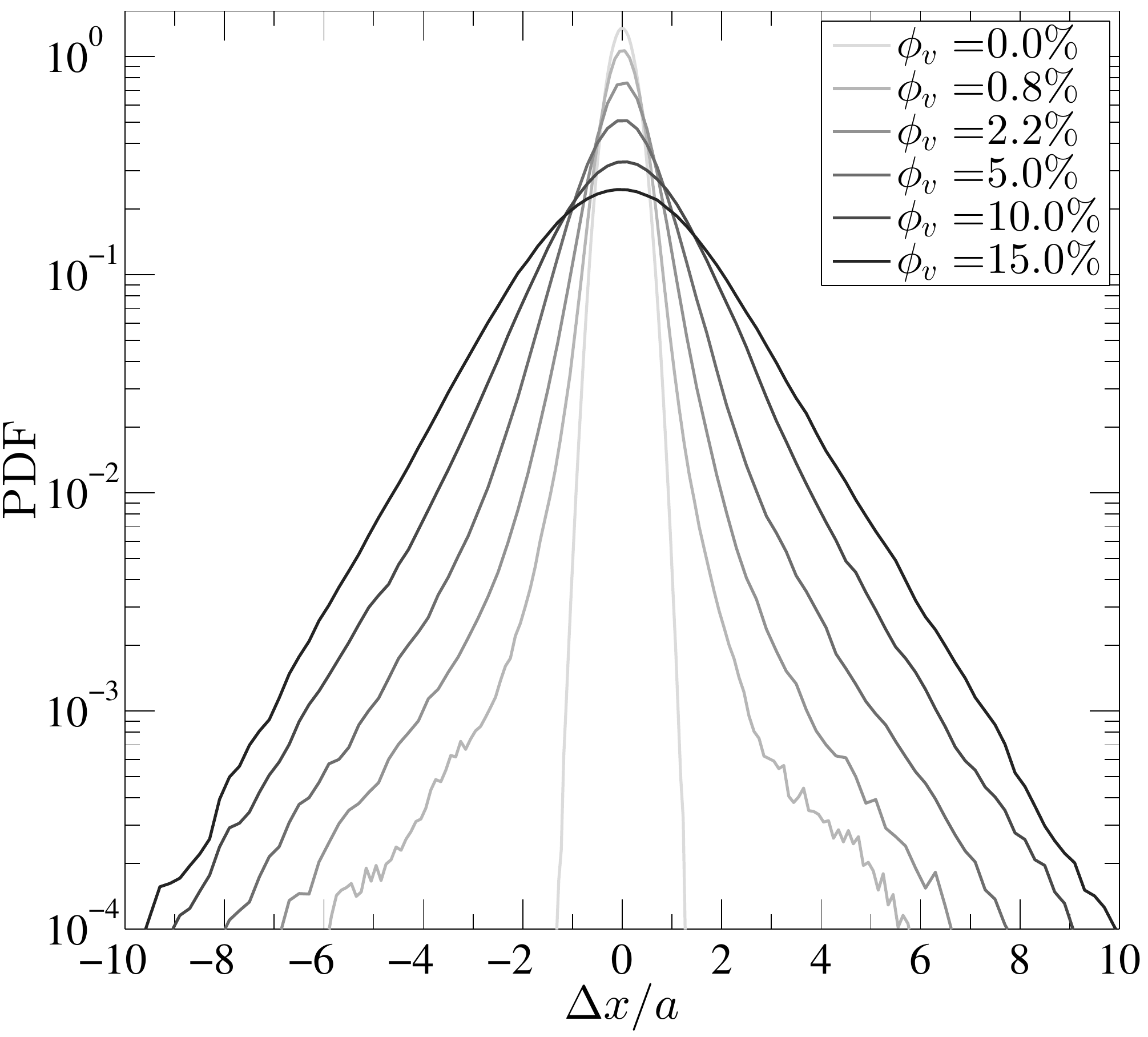}}
\subfloat[$\Delta t = 2s$]{\label{fig:PDF_disp_concentrated_DT_2} \includegraphics[width=0.5\columnwidth]{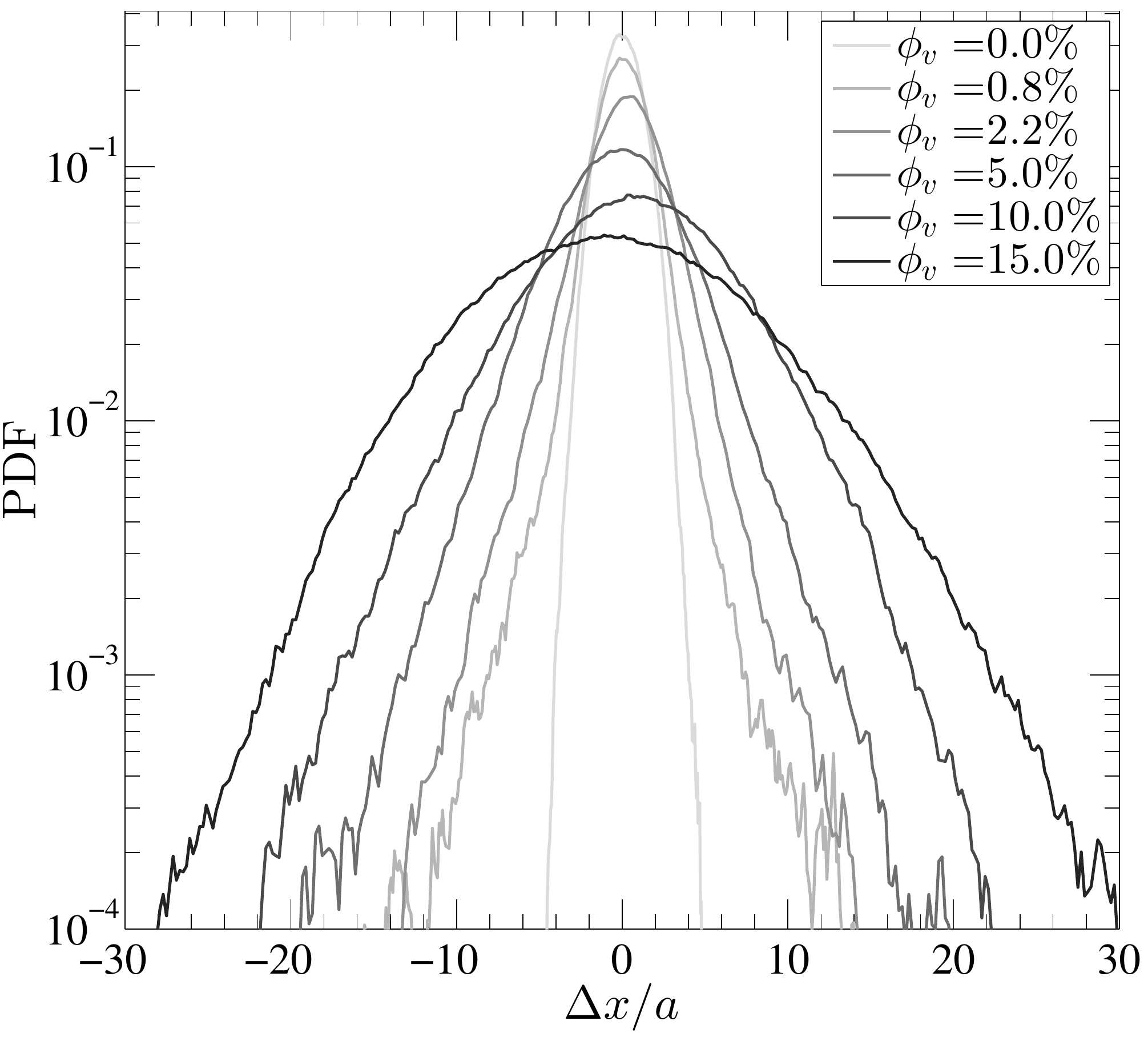}}
\end{centering}
\caption{\label{PDF_disp_concentrated} PDF for tracer displacements at times $\Delta t = 0.3s$ (a), and $\Delta t = 2s$ (b), for $\phi_v = 0 - 15\%$.
}
\end{figure}

In this section, we perform simulations on more concentrated suspensions using the same value for the squirming parameter $\beta=0.5$. Figure \ref{PDF_disp_concentrated} shows the time-dependent PDF for tracer displacements $P(\Delta x/a, \Delta t)$ at times $\Delta t = 0.3s$ and $\Delta t = 2s$, and for volume fractions $\phi_v = 0-15\%$.  For all volume fractions, we observe that the PDF tends towards a Gaussian at long times, however, the convergence rate increases with volume fraction.  Using an analysis based on singular flows fields, \cite{Zaid2011} state that the tracer displacement distribution should only be Gaussian for $\phi_v>25\%$ if the flows generated by the swimmers decay like $r^{-n}$ with $n\geq2$.  The appearance of a Gaussian, however, agrees with the theoretical predictions of \cite{Thiffeault2014}, as well as the experimental observations of \cite{Kurtuldu2011} that observed an increase in Gaussianity with concentration.  This, perhaps, highlights the importance of resolving the finite-size of both swimmers and tracers to the observation of a Gaussian distribution of displacements.  In addition, we note a shift of the mean at long times for high volume fraction.  This is attributed to the onset of polar ordering (\cite{Delmotte_2015a}) that is typically observed in periodic squirmer suspensions. 

\begin{figure}
\centering
\subfloat[Effective diffusion coefficient of tracers. \symbol{\solid}{\bigcircle}{20}{0}{black}{black}: simulations. 
\symbol{\dashed}{}{0}{0}{black}{black}: linear fit for the dilute regime $\phi_v = 0-2.2\%$. 
\symbol{}{\ssquareb}{0}{0}{black}{blue}: data from \cite{Leptos2009}.]{\label{fig:Deff_concentrated} \includegraphics[width=0.45\columnwidth]{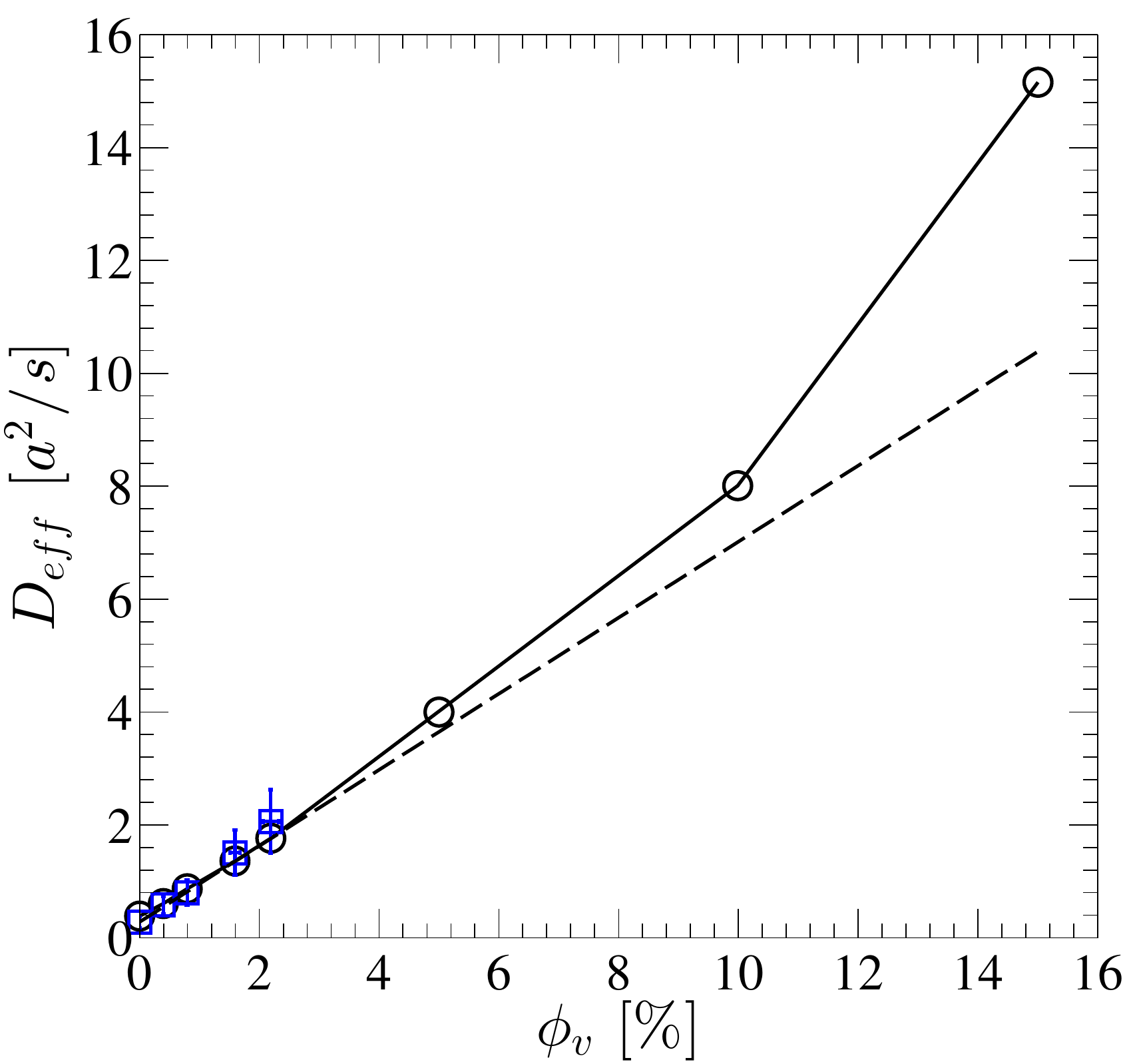}}
\hspace{0.5cm}
\subfloat[Mean-squared displacement of swimmers. Inset: zoom at long times to show the decrease of the slope with the concentration.]{\label{fig:MSD_swimmers} \includegraphics[width=0.45\columnwidth]{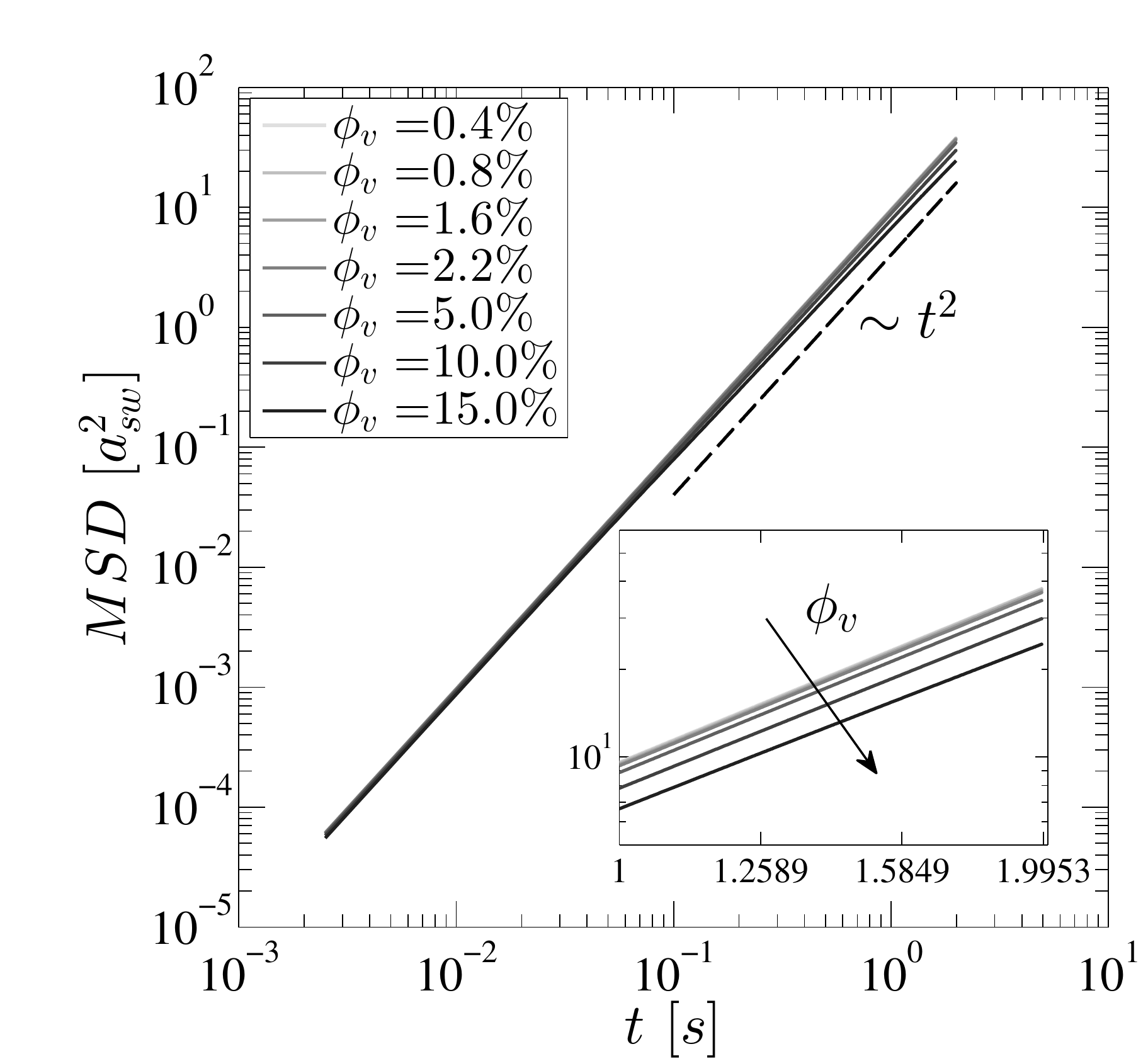}}
\caption{Effective diffusion coefficient of tracers (a) and mean-squared displacement of swimmers (b) with $\phi_v = 0 - 15\%$.
}
\end{figure}

As in the experiments of \cite{Kurtuldu2011} and \cite{Kasyap2014}, our simulations show (Figure \ref{fig:Deff_concentrated}) the linear scaling $(D_{eff}-D_0)/\phi_v =$ const. breaks down for $\phi_v>2.2\%$, though we do not observe a clear power-law as in \cite{Kurtuldu2011}.  Additionally, our simulations yield $D_{eff}/D_0 = 11 - 22$ for $\phi_v = 5-10\%$, while \cite{Kurtuldu2011} measured $D_{eff}/D_0 \simeq 900$ for $\phi_v = 7\%$.  The large difference in enhanced diffusion can be due to our simulations being three-dimensional, while \cite{Kurtuldu2011} carried out experiments into a two-dimensional film with thickness on the order of the cells' diameter ($H\simeq 15 \pm 5 \mu m$), resulting in a slower decay of the flows induced by the swimming cells, as well as an increased frequency of swimmer-tracer collisions.  To quantify the effect of swimmer concentration on the swimmers' motion, we examine the swimmer mean-squared displacement as a function of concentration, as shown in Figure \ref{fig:MSD_swimmers}.  While we see a slight decrease in the slope as $\phi_v$ increases, we observe that the swimmers are ballistic for all concentrations.  Longer simulation times would be required to reach a diffusive regime in order to determine swimmer effective diffusivity $D_{eff,sw}$, as done in \cite{Ishikawa2010}.  Our simulations, therefore, correspond to a different regime than that explored in \cite{Ishikawa2010} where both swimmers and tracers were diffusive.  We note, however, that the values $|\beta| > 1$ used in \cite{Ishikawa2010} are greater than the values used here to match experimental data and higher values of $\beta$ do lead to the onset of diffusive behaviour for the swimmers at earlier times.  

\begin{figure}
\begin{centering}
\subfloat[$\phi_v = 2.2\%$]{\label{fig:RDF_phi_2_2} \includegraphics[width=0.5\columnwidth]{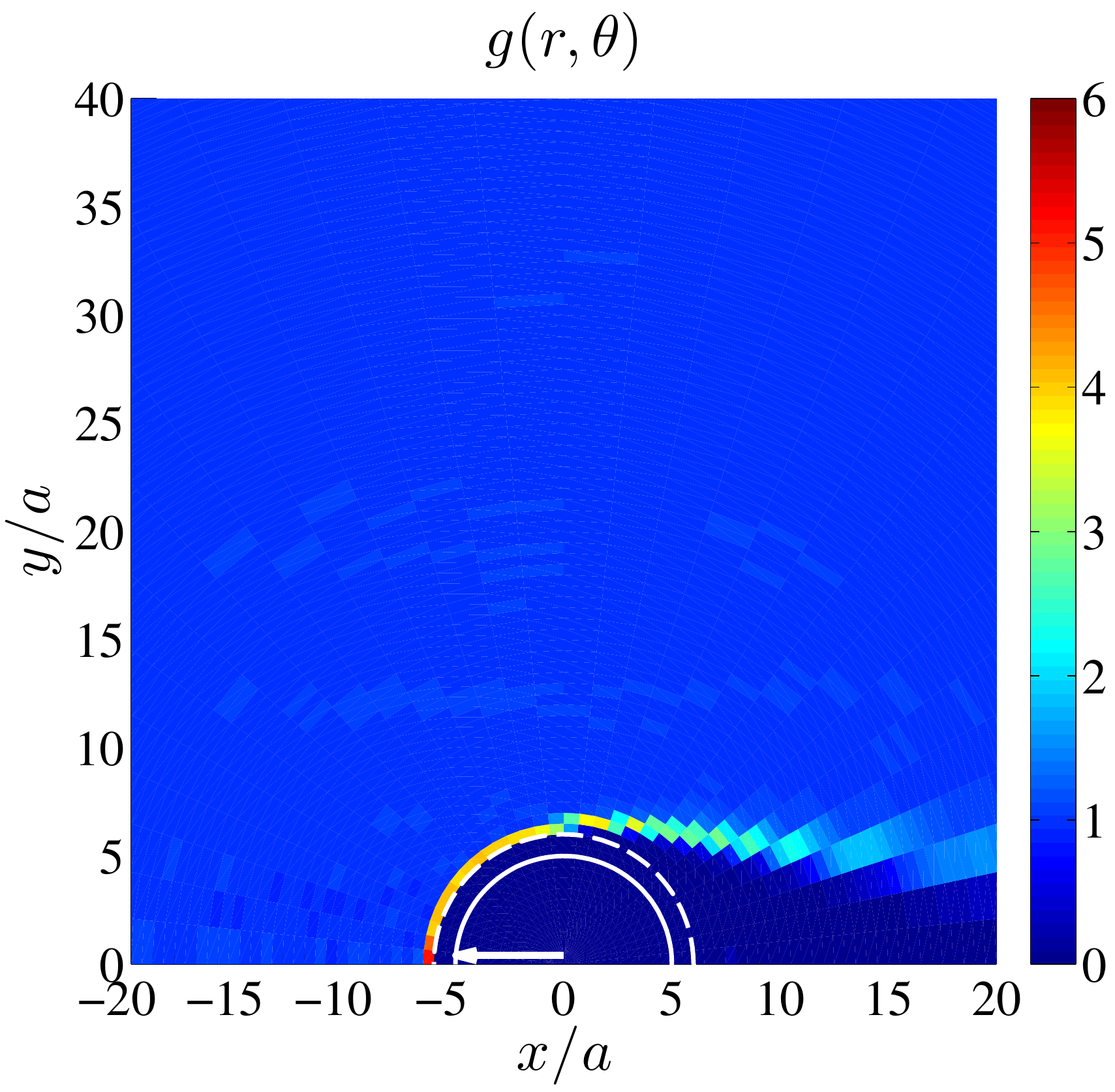}}
\subfloat[$\phi_v = 15\%$]{\label{fig:RDF_phi_15} \includegraphics[width=0.5\columnwidth]{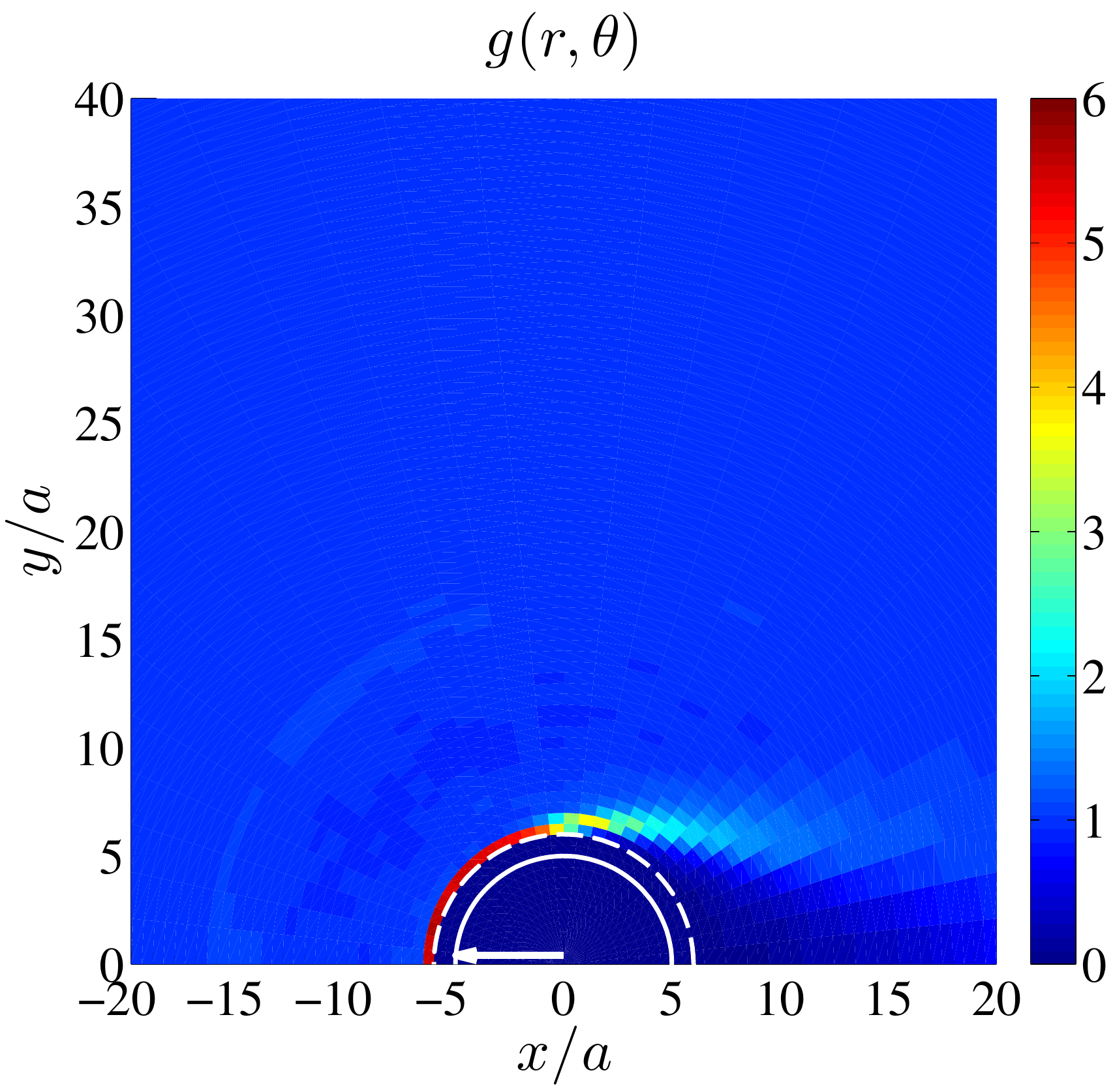}}
\end{centering}
\caption{\label{fig:RDF_phi} Squirmer-tracer pair distribution function $g(r,\theta)$ for $\beta = 0.5$. The white solid line represent the swimmer's radius. The white dashed line corresponds to the excluded volume region. The white arrow is the swimming direction.
}
\end{figure}

While for each concentration the swimmers' motion is ballistic, we do observe clear differences in the tracer distribution as the swimmer concentration increases.  Figure \ref{fig:RDF_phi} shows the squirmer-tracer pair distribution function $g(r,\theta)$ for $\phi_v = 2.2$ and $15\%$.  The distribution function $g(r,\theta)$ corresponds to the likelihood, relative to a uniform distribution, of finding a tracer at a distance $r$ from the swimmer centre of mass and with its relative position vector forming an angle $\theta$ with the swimmer orientation.  At higher concentrations, there is a much higher probability of finding a tracer directly in front of the swimmer, while the probability of finding tracers aft of the swimmer is reduced.  This suggests that accurately representing the details of swimmer-tracer interactions for tracers directly in front of the swimmer is important in reproducing the large displacements.  In addition, the increased localisation of tracer particles in the vicinity of the swimmers that occurs at higher swimmer concentrations could lead to increased rates of nutrient uptake by the population.

\section{Discussion and conclusion}
\label{sec:Discussion}

In this study, we performed simulations of tracer motion in dilute and semidilute suspensions of swimming particles.  Our model includes the coupled effects of the finite tracer size relative to that of the swimmers, particle Brownian motion satisfying the fluctuation-dissipation theorem, and the flow disturbances induced by the swimmers using the steady squirmer model.  For dilute suspensions, our simulations reproduce quantitatively the tracer displacement distribution and tracer diffusivity measured experimentally by \cite{Leptos2009}, as well as those predicted theoretically by \cite{Thiffeault2014}.  We demonstrate that the non-Gaussian tails of the tracer displacement distribution are linked to collisions between the swimmers and tracers, while the width of the Gaussian core is set by the many-body hydrodynamic interactions between the tracers and swimmers, as well as latent tracer diffusion.  We also show that time-dependence of the squirmer modes has little effect on the distribution.  In addition, our results demonstrate that at longer observation times, the number of swimmer-tracer collisions increase leading to the disappearance of the non-Gaussian power law tails and a Gaussian tracer displacement distribution.  This corresponds with the predictions of \cite{Thiffeault2014}.  In the semi-dilute regime, the swimmer-tracer collision rate increases and we observe a faster convergence to a Gaussian distribution of displacements.  Additionally, the simulations yield a nonlinear dependence of the effective diffusion coefficient on $\phi_v$ though the enhancement observed was much more modest than that measured (\cite{Kurtuldu2011}) in the suspensions confined to a thin film.  In the nonlinear regime, we found that at the simulation timescales, the swimmers were still behaving ballistically, but there was a notable increase in the likelihood of finding tracers directly in front of the swimmers.  

A particularly interesting property of this system is the importance of near contact swimmer-tracer collisions, even in dilute systems, as they lead to rare, but very large displacements.  Even though the squirmer model provides a reasonable description of the far-field flow of many microorganisms, it is the near-field details that matter when resolving the collisions (\cite{Pushkin2013b}).  Thus, at longer times when there are sufficiently many collision events, the effective tracer diffusion coefficient will be determined almost exclusively by the collisions.  The details of how the organisms generate flow and possibly deform their surfaces or move their flagella are then crucial in quantifying the effective diffusion (\cite{Polin2016}).  This could have important implications in predator-prey interactions, making only certain locomotion strategies viable for particle capture.  For example, a swimmer with a rigid front surface could possibly be more effective at carrying along small particles as it swims, allowing more time for them to be ingested.  In addition, the importance of near-field interactions on particle diffusion may certainly impact the design of synthetic artificial swimmers or other colloidal active particles for tracer mixing in microfluidic devices.  Exploring the impact of different near-field swimmer-tracer interactions for swimmers that induce the same far-field flow, as well as how heterogeneity throughout the swimmer population affect tracer transport are areas of interest for future investigation.

In bacterial suspensions, the nonlinear dependence of the effective tracer diffusion coefficient on swimmer volume fraction has been attributed (\cite{Kasyap2014}) to the onset of large-scale collective motion in the suspension.  In squirmer suspensions, especially at semi-dilute concentrations, the long-time dispersion properties might well be affected by polar ordering (\cite{Delmotte_2015a}), which may lead to net tracer transport along a particular axis, possibly aligned with the mean swimming direction.  

Finally, it would be interesting to understand the impact of confinement on tracer diffusivity, especially with regard to the thin film experiments of \cite{Kurtuldu2011}.  Due to the stress-free boundaries, the hydrodynamic interactions are longer ranged than in bulk and may therefore enhance tracer diffusivity more than in unconfined suspensions.  Additionally, we expect the swimmer-tracer collision rate to increase in confined systems which should further increase long time tracer diffusivity.  Our numerical framework can simulate thin film geometries with stress-free boundary conditions while retaining Brownian motion and hydrodynamic interactions (\cite{Delmotte_2015b}).  These simulations may provide some insight into the role of confinement and boundary condtions on tracer transport and form the basis of future studies.


\acknowledgement{
This work is developed within the MOTIMO ANR Project. Simulations were performed on the Calmip supercomputing mesocenter. We thank INPT for funding the international collaboration between IMFT and Imperial College (grant no. SMI 2014).  EEK gratefully acknowledges support from EPSRC under grant EP/P013651/1.  The authors also thank the many members of the COST Action MP1305 on Flowing Matter for fruitful discussions.}

\bibliographystyle{apalike}
\bibliography{artbib}

\end{document}